\documentclass[a4paper,11pt]{article}
\usepackage{jheppub}

\arxivnumber{2611.04802}

\title{\boldmath Study of enhancing the signal-background separation for the Higgs boson production in association with a single top-quark in $pp$ collisions at the LHC}

\author[a,b]{R.T.~Azimzade}
\author[a]{A.E.~Dadashova}
\author[a]{I.R.~Boyko}
\author[a]{V.V.~Lyubushkin}
\author[a,b]{N.A.~Huseynov}

\affiliation[a]{Dzhelepov Laboratory of Nuclear Problems, Joint Institute for Nuclear Research,\\
Dubna, Moscow Region, Russia}
\affiliation[b]{Institute of Physics, Ministry of Science and Education of Azerbaijan,\\
Baku, Azerbaijan}

\emailAdd{rustam.azimzade@cern.ch}

\abstract{In this paper, a multivariate analysis based approach for enhancing the extraction of the rare Standard Model (SM) process being studied - Higgs boson production in association with a single top quark ($tH$) - from the prevailing background processes is presented. The analysis was performed using a Monte Carlo generated dataset simulating the outcomes of proton-proton collisions at the Large Hadron Collider (LHC). A deep fully connected neural network was developed for signal-background discrimination, incorporating a masking mechanism designed to suppress the contribution of input variables associated with objects that were not reconstructed in a given event. The performance of the masked architecture was evaluated on independent Monte Carlo samples. The results of the proposed methodology are compared with a conventional cut-based approach in terms of the statistical significance of the $tH$ production channel. The developed framework was tested using a simulated dataset corresponding approximately to the combined statistical samples of LHC Runs 1, 2, and 3.}

\begin{document}
\maketitle
\flushbottom

\section{Introduction}
\label{sec:introduction}

The Standard Model (SM) \cite{10.1103/physrevlett.19.1264, 10.1016/0029-5582(61)90469-2, 10.1142/9789812795915_0034} is the most comprehensive theory describing the fundamental interactions of elementary particles: quarks, leptons, and bosons. The construction of this theory was completed by discovering the Higgs boson \cite{10.1103/PhysRevLett.13.321, 10.1103/PhysRevLett.13.508} both in the ATLAS and CMS experiments \cite{10.1016/j.physletb.2012.08.020,10.1016/j.physletb.2012.08.021} in 2012. However, deviations between theoretical predictions and experimental data continue to promote more precise measurements of its parameters. Moreover, unresolved issues drive scientists to search for new physical phenomena beyond the Standard Model (BSM).

Precise measurement of the Higgs boson parameters is of particular interest in elementary particle physics. Despite its experimental confirmation, some of its properties, such as the possible presence of CP symmetry violation \cite{10.1103/PhysRevLett.76.4468, 10.1007/jhep04(2014)004, 10.1142/S0217751X15501560} or the detailed structure of the Higgs potential, remain not fully explored. Specifically, the Higgs potential is sensitive to radiative corrections \cite{10.1007/jhep08(2012)098, 10.1007/jhep12(2013)089}, in which the top Yukawa coupling provides a dominant contribution, thereby affecting the stability of the electroweak (EW) vacuum \cite{10.1088/0954-3899/41/11/113001}. In this context, a more detailed investigation of the Higgs boson’s interactions holds significant potential for uncovering signs of ``new physics'' \cite{10.1007/jhep12(2013)089} beyond the Standard Model.

\begin{figure}[htbp]
\centering
\includegraphics[width=.47\linewidth]{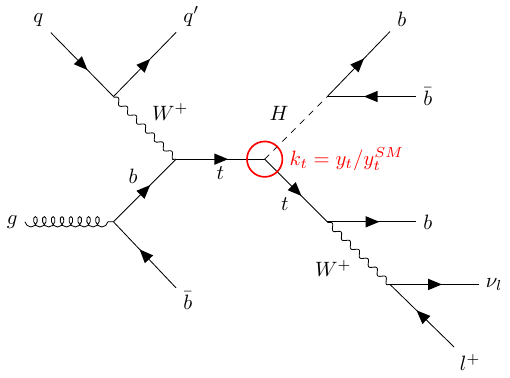}
\qquad
\includegraphics[width=.4\linewidth]{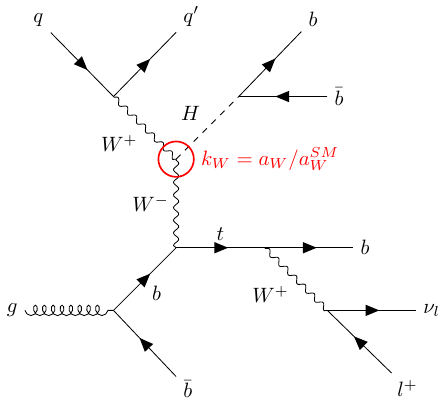}
\caption{Feynman diagrams of leading contributions for signal Higgs boson production in association with a single top-quark, followed by the Higgs boson decay into a $b\bar{b}$ pair and top-quark decay via leptonic mode.\label{fig:tH_channels}}
\end{figure}

To date, Higgs boson production has been observed in four channels: two production channels alongside other elementary particles (Higgs strahlung and Top fusion), and two where a single Higgs boson is produced alone (Gluon-Gluon Fusion and Vector Boson Fusion). Among all other hypothetical Higgs boson production channels, the most intriguing is its production in association with a single top quark ($tH$) \cite{10.1103/physrevd.64.094023, 10.1140/epjc/s10052-015-3475-9, 10.1051/epjconf/201920104003}. The first two leading-order perturbation theory Feynman diagrams for ($tH$) process are shown in figure~\ref{fig:tH_channels}.

The magnitude of the top-Higgs Yukawa coupling $y_t = |y_t|e^{i\phi_t}$ has already been indirectly constrained with a precision of approximately $\mathcal{O}(10\%)$ at the LHC~\cite{10.1038/s41586-022-04893-w,10.1140/epjc/s10052-019-6909-y}. Direct measurements of $|y_t|$ are provided by the $t\bar{t}H$ production process, whereas the $tH$ process is directly sensitive to the relative complex phase $\phi_t$ between the $ttH$ and $WWH$ couplings through the interference of the corresponding production amplitudes~\cite{10.1103/PhysRevD.63.014018, 10.1103/PhysRevD.81.034020, 10.1007/jhep01(2013)088, 10.1007/jhep05(2013)022, 10.1007/jhep05(2014)062}. In the Standard Model, this interference is destructive, suppressing the $tH$ production rate, while a change in the relative phase can lead to constructive interference. Consequently, the $tH$ process provides a unique probe of the relative phase of the top-Higgs Yukawa coupling, while statistically significant evidence for its production has yet to be established.

\begin{figure}[htbp]
\centering
\includegraphics[width=.5\linewidth]{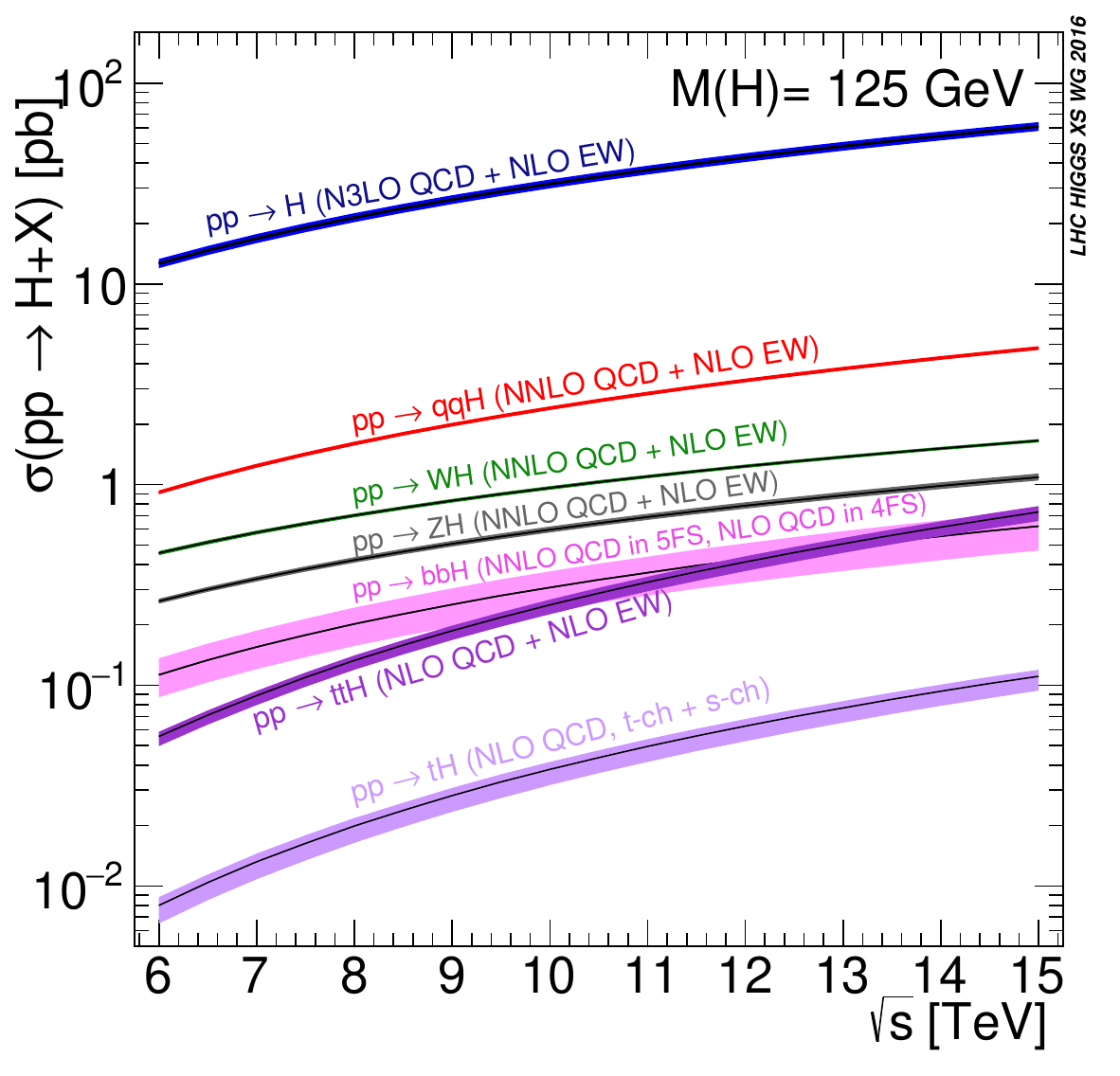}
\caption{Dependence of cross-section for various  Standard Model Higgs boson productions on the center-of-mass energy $\sqrt{s}$ of proton-proton collisions \cite{CERN_HiggsXSBR}.\label{fig:all_channels}}
\end{figure}

Being of great interest for experimental confirmation and study \cite{10.1103/physrevd.64.094023}, this production channel is particularly challenging to detect because the Standard Model predicts the smallest production cross-section for $tH$ (see figure~\ref{fig:all_channels}). In this paper, the potential for observing the $tH$ production process is studied in proton-proton collisions at a centre-of-mass energy of $\sqrt{s}=13.6~\mathrm{TeV}$, assuming an integrated luminosity of $L=500~\mathrm{fb}^{-1}$. Destructive interference leads to the suppression of the total interaction cross-section to $69.66~\mathrm{fb}$ at $\sqrt{s}=13.6~\mathrm{TeV}$ \cite{10.1007/jhep01(2013)088}, resulting in an extremely small signal yield compared with the background.

Background processes comprise a multitude of possible events occurring in proton–proton collisions, distinct from the Higgs boson production in association with a single top quark ($tH$), which is the primary focus of this study. A large fraction of these background contributions can be suppressed by exploiting the presence of specific physics objects in the final-state signature. The most reasonable approach to extract the $tH$ events is to select the decay modes which ensure simultaneously the highest signal statistics and suppression of the majority of background channels. For this reason, when defining the signal process signature, special attention was given to the top quark decay via the semileptonic mode $(t\to{bl\nu_l})$, which accounts for $33.2\%$ of all top quark decays, and the Higgs boson decay through its dominant mode $(H\to{b}\bar{b})$, which occurs in $58\%$ of cases. The signal process is characterized by the presence of at least three $b$-jets, a charged lepton with high transverse momentum and significant missing transverse energy carried away by a neutrino. An additional ``forward jet'', originating from the spectator quark $q^{\prime}$, is also characteristic of the $tH$ production topology (see figure~\ref{fig:tH_channels}) \cite{10.1134/s1063778822020041}.

However, there exists a class of background processes whose final-state signatures exactly mimic that of the signal. Consequently, further discrimination based solely on object-level selection becomes insufficient, necessitating the use of multidimensional kinematic information to enhance the signal-to-background separation. This issue is further exacerbated by the fact that at $13.6~\mathrm{TeV}$ the $(pp\to tH)$ production cross-section is four orders of magnitude smaller than the main background process ($pp\to t\bar{t}$) with $811~\mathrm{pb}$ of cross-section \cite{10.1007/jhep04(2019)046,10.1140/epjc/s10052-019-6863-8}. As a result, signal events are vastly outnumbered by the dominant background processes, making their extraction from the prevailing and kinematically similar background particularly challenging.

The outcome of each collision event is characterized by a finite set of kinematic observables that describe the corresponding reconstructed physical objects in the final state. Each event can therefore be represented by a feature vector $\mathbf{x} = (x_1, x_2, \dots, x_d)$, where $d$ denotes the number of variables used. The collection of such features forms a $d$-dimensional phase space in which each event corresponds to a single point $\mathbf{x}$. Signal and background events are distributed in this space according to the probability density functions $p_s(\mathbf{x})$ and $p_b(\mathbf{x})$, respectively.

Intuitively, the so-called signal region can be identified in those regions of the feature space where the probability of observing a signal event exceeds that of observing a background event, i.e. $p_s(\mathbf{x})>p_b(\mathbf{x})$. In general, the signal and background regions do not necessarily form single connected domains with clear boundaries. Instead, since there may exist regions in which the probability densities of the corresponding distributions coincide, the signal-enriched regions may overlap with the background-enriched regions. Furthermore, the complex nonlinear structure of the signal probability density $p_s(\mathbf{x})$ in phase space significantly limits the effectiveness of conventional cut-based approaches for optimal signal region definition beyond a certain stage of the event selection procedure.

Therefore, the signal–background classification problem is inherently statistical in nature and consists in constructing a decision rule that minimizes the probability of classification error. From the perspective of statistical decision theory, the optimal classification rule is determined by the Neyman–Pearson lemma. According to this lemma, the most powerful test for discriminating between the signal and background hypotheses is based on the likelihood ratio \cite{10.1098/rsta.1933.0009}:

\begin{equation}
\Lambda(\mathbf{x})=\frac{p_s(\mathbf{x})}{p_b(\mathbf{x})}
\end{equation}

Event classification is performed by introducing a threshold value $\tau$, which defines a separating hypersurface in the feature space corresponding to the boundary $\Lambda(\mathbf{x})=\tau$. In this case, the decision rule can be written as:

\begin{equation}
\Lambda(\mathbf{x})
\begin{cases}
> \tau, & \text{signal region}, \\
< \tau, & \text{background region}.
\end{cases}
\end{equation}

The choice of the threshold value $\tau$ is determined according to a selected statistical criterion in order to maximize the sensitivity to the signal while minimizing the statistical uncertainty.

In practice, obtaining explicit analytical expressions for the probability density functions $p_s(\mathbf{x})$ and $p_b(\mathbf{x})$ is generally infeasible, since their evaluation involves multidimensional phase-space integration of matrix elements including higher-order QCD contributions. Moreover, due to the similarity of the final state signatures of signal and some background events in the collider, the true origin of each recorded event (signal or background) is unknown. This precludes the direct use of experimental data for reconstructing these distributions. Instead, in order to obtain labeled datasets suitable for further analysis, specialized event generators are employed, which incorporate up-to-date theoretical models of elementary particle physics. These labeled datasets provide an approximate representation of the real distributions in the feature space.

In this study, we present the development of a methodology aimed at improving the efficiency of identifying the rare process $(pp\to tHbq)$. The analysis is performed using Monte-Carlo generated datasets that include both signal and background processes. 

Achieving the highest possible signal region determination accuracy will facilitate reaching the signal significance level sufficient for an observation and confirmation of the $tH$ production channel at the earliest possible stage. Furthermore, in future measurements of various physical properties in this channel, achieving the best possible selection accuracy will also improve the precision of estimating the corresponding physical parameters.

\section{Monte-Carlo samples}
\label{sec:generation}

\begin{figure}[htbp]
\centering
\includegraphics[width=.25\linewidth]{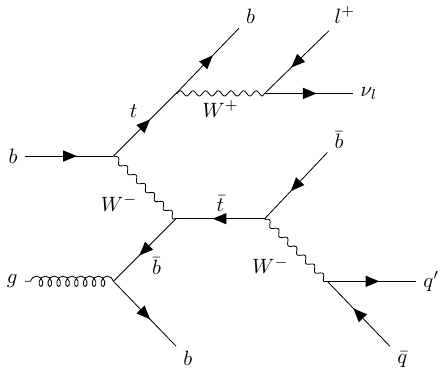}
\qquad
\includegraphics[width=.25\linewidth]{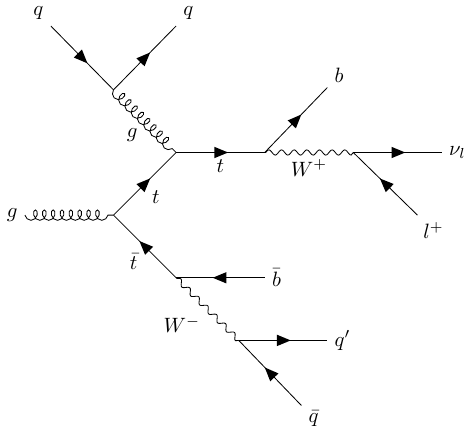}
\qquad
\includegraphics[width=.25\linewidth]{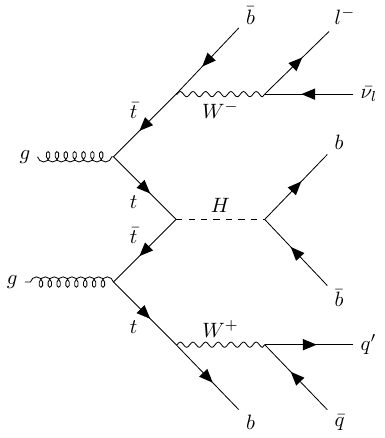}
\qquad
\includegraphics[width=.25\linewidth]{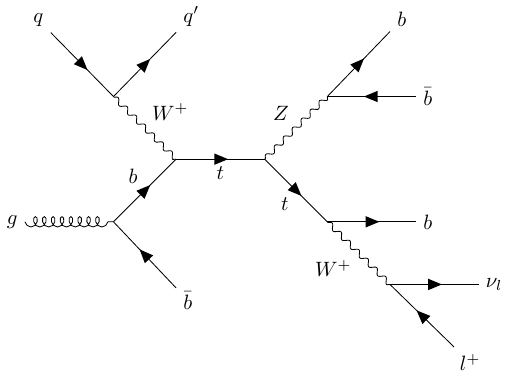}
\qquad
\includegraphics[width=.25\linewidth]{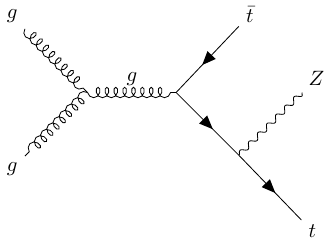}
\qquad
\includegraphics[width=.25\linewidth]{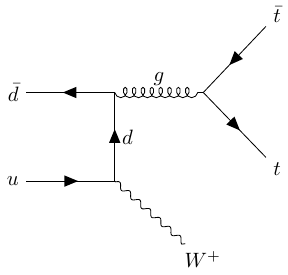}
\caption{Leading channels for each background process. Where each Feynman diagram shows the first order of perturbation theory for $t\bar{t}$, $t\bar{t}b\bar{b}$, $t\bar{t}H$, $tZbq$, $t\bar{t}Z$, and $t\bar{t}W$ processes, respectively.\label{fig:background}}
\end{figure}

The signal and most background processes ($t\bar{t}H$~\cite{10.1016/j.physletb.2018.07.035}, $t\bar{t}Z$~\cite{10.1007/jhep03(2020)056}, $tZbq$~\cite{10.1016/j.physletb.2018.03.023}, $t\bar{t}W$~\cite{10.1103/physrevlett.131.231901}, $t\bar{t}b\bar{b}$~\cite{10.3390/universe9050242}) were simulated using the MadGraph5\_aMC@NLO~\cite{10.1007/jhep07(2014)079} Monte Carlo (MC) generator, while parton showering and hadronization were modeled with the Pythia8~\cite{2203.11601}. The leading-order Feynman diagrams for all considered background processes are presented in figure~\ref{fig:background}. The main background process ($t\bar{t}$) was simulated using POWHEG~\cite{10.1007/jhep02(2024)023} and Herwig7~\cite{10.1140/epjc/s10052-016-4018-8} as the parton shower generator. The matrix elements for signal and background processes with the largest cross-section ($t\bar{t}$, $t\bar{t}H$) were calculated in the next-to-leading order of perturbation theory. Jet clustering was performed using the FastJet package~\cite{10.1140/epjc/s10052-012-1896-2}. The following kinematic selection criteria have been applied to isolate the signal reactions $pp\rightarrow tH, t\rightarrow b l\nu_l, H\rightarrow b\bar{b}$ (see figure~\ref{fig:tH_channels} and section~\ref{sec:introduction}) \cite{10.1134/s1063778822020041}:

\begin{itemize}
    \item Transverse momentum ($p_{T}^{\text{lead lep}}$) for the leading lepton (the charged lepton with the highest $p_{T}$) of at least $25 \,\text{GeV/c}$. This restriction corresponds to one of the triggers for recording a charged lepton event in the ATLAS experiment.
    \item $|\eta^{lead\text{ }lep}| < 2.7$, where $\eta^{lead\text{ }lep}$ is the pseudorapidity of the leading lepton; accounting for the geometry of the ATLAS and CMS detectors.
    \item $p_{T}^{\nu, reco} \geq 10$ GeV, where $p_{T}^{\nu,reco}$ is the transverse momentum of the reconstructed neutrino; the missing transverse energy is reconstructed in the ATLAS and CMS experiments with a resolution of around 10 GeV.
    \item $|\eta^{b-jet}| < 2.4$, where $\eta^{b-jet}$ is the pseudorapidity of the b-jets; a limit roughly corresponding to the acceptance of the ATLAS~\cite{10.1140/epjc/s10052-017-4852-3, 10.1088/1748-0221/19/06/p06029} and CMS~\cite{10.1088/1748-0221/12/01/p01020, 10.1088/1748-0221/19/11/p11021, 10.1088/1748-0221/19/05/p05064} tracking systems.
\end{itemize}

Table~\ref{tab:yields} provides information on the number of events simulated and selected for inclusion in the database for signal and
background processes.

\begin{table}[htbp]
\centering
    \begin{tabular}{l|c|c|c|c|c}
    \hline
      Process & ME order & $\sigma~(\mathrm{fb})$ & $N_{\text{generated}}$ & $N_{\text{preselected}}$ & Preselection yield \\
       \hline
       $tHbq$ & NLO & $13.4$ & $8.78 \times 10^6$ & $1.08 \times 10^6$ & $822$ \\
       $t\bar{t}$ & NLO & $4.05 \times 10^5$ & $3.00 \times 10^7$ & $7.67 \times 10^4$ & $5.18 \times 10^5$ \\
       $t\bar{t}b\bar{b}$ & LO & $4.81 \times 10^3$ & $6.60 \times 10^7$ & $4.07 \times 10^6$ & $1.48 \times 10^5$ \\
       $t\bar{t}H$ & NLO & $129$ & $2.75 \times 10^6$ & $5.59 \times 10^5$ & $1.31 \times 10^4$ \\
       $t\bar{t}Z$ & LO & $34.1$ & $2.20 \times 10^7$ & $3.81 \times 10^6$ & $2.95 \times 10^3$ \\
       $tZbq$ & LO & $31.4$ & $2.20 \times 10^7$ & $2.31 \times 10^6$ & $1.65 \times 10^3$ \\
       $t\bar{t}W$ & LO & $212$ & $2.20 \times 10^7$ & $1.64 \times 10^5$ & $792$ \\
       \hline
       Total background &  &  & $1.65 \times 10^8$ & $1.10 \times 10^7$ & $6.84 \times 10^5$ \\
       \hline
    \end{tabular}
    \caption{Statistics of simulated and selected events in this study. $N_{\text{generated}}$ is the number of generated events; $\sigma$ is the process cross-section; $N_{\text{preselected}}$ is the number of generated events passing the selection described above; The preselection yield is the expected number of events after the preliminary cuts, normalized to an integrated luminosity of $500~\text{fb}^{-1}$ using the process cross-sections and the selection efficiencies estimated from simulation.\label{tab:yields}}
\end{table}

The neutrino transverse momentum ($E_{\mathrm{T}}^{\mathrm{miss}}$) was reconstructed as the vector opposite to the total transverse momentum of all jets and charged leptons. To determine the longitudinal component of the neutrino momentum, the invariant mass of the leading lepton and the neutrino was set equal to the nominal mass of the W-boson ($m_{W} = 80.35$ GeV).

This constraint leads to two solutions:

\begin{equation}
    p^{z,\pm}_{\nu, reco} = \frac{1}{(p_{T}^{lead\text{ }lep})^{2}}\left( \mu\cdot p^{z}_{lead\text{ }lep} \right. \\
   \left. \pm E_{lead\text{ }lep}\sqrt{ \mu^{2} - (p_{T}^{lead\text{ }lep}E_{T}^{miss})^{2}}\right),
\end{equation}
\noindent
where $\mu = \frac{m_{W}^{2}}{2} + p^{x}_{lead\text{ }lep}E_{T}^{miss,x} + p^{y}_{lead\text{ }lep}E_{T}^{miss,y}$.

The jet clustering was performed using the anti-$k_T$ algorithm \cite{10.1088/1126-6708/2008/04/063} with the following parameters: minimum jet transverse momentum $p_{T}^{min} = 20~\mathrm{GeV}$, jet radius parameter $R = 0.5$, and pseudorapidity range for all jets $|\eta| < 5.0$. 

A jet was tagged as a "b-jet" only if the distance between the b-quark momentum and the jet momentum satisfied $\Delta R < 0.1$ and the ratio of the jet energy to the b-quark energy was in the range $0.5 \leq \frac{E_{j}}{E_{b}} \leq 1.5$. 

The main background process for the studied Higgs boson production channel is $pp\rightarrow t\bar{t}$; given the requirement of one charged lepton, the decay channel takes the form $pp\rightarrow t\bar{t}\rightarrow (bl\nu_l)(bqq')$. To first approximation, the background process can be suppressed by requiring at least three b-jets in the event. Due to the limited precision of experimental detectors, a c-quark jet can be misidentified as a b-jet with a probability of about $12\%$ in the ATLAS \cite{10.1088/1748-0221/11/04/p04008} and CMS \cite{10.1088/1748-0221/13/05/p05011} experiments. Jets from light quarks ($u$, $d$, $s$) are misidentified as b-jets with a probability of about $0.3\%$ \cite{10.1088/1748-0221/11/04/p04008, 10.1088/1748-0221/13/05/p05011}. This effect was taken into account by assigning a $b$-tag to random jets from $c$ and ($u$, $d$, $s$) quarks with corresponding probability when simulating the $pp\rightarrow t\bar{t}$ process.

Event reconstruction requires an assignment of every $b$-jet to the decay of either top quark or Higgs boson. This is performed by selecting the combination that provides the best agreement between the reconstructed masses of the top quark and the Higgs boson and their nominal values. The top quark mass is reconstructed from the leading lepton, neutrino, and a b-jet, while the Higgs boson mass is reconstructed from the remaining two b-jets. In the case when four b-jets were reconstructed in the event, three of them with the highest $p_T$ were selected for the reconstruction of the top-quark and Higgs boson masses.

\begin{figure}[htbp]
\centering
\includegraphics[width=.4\linewidth]{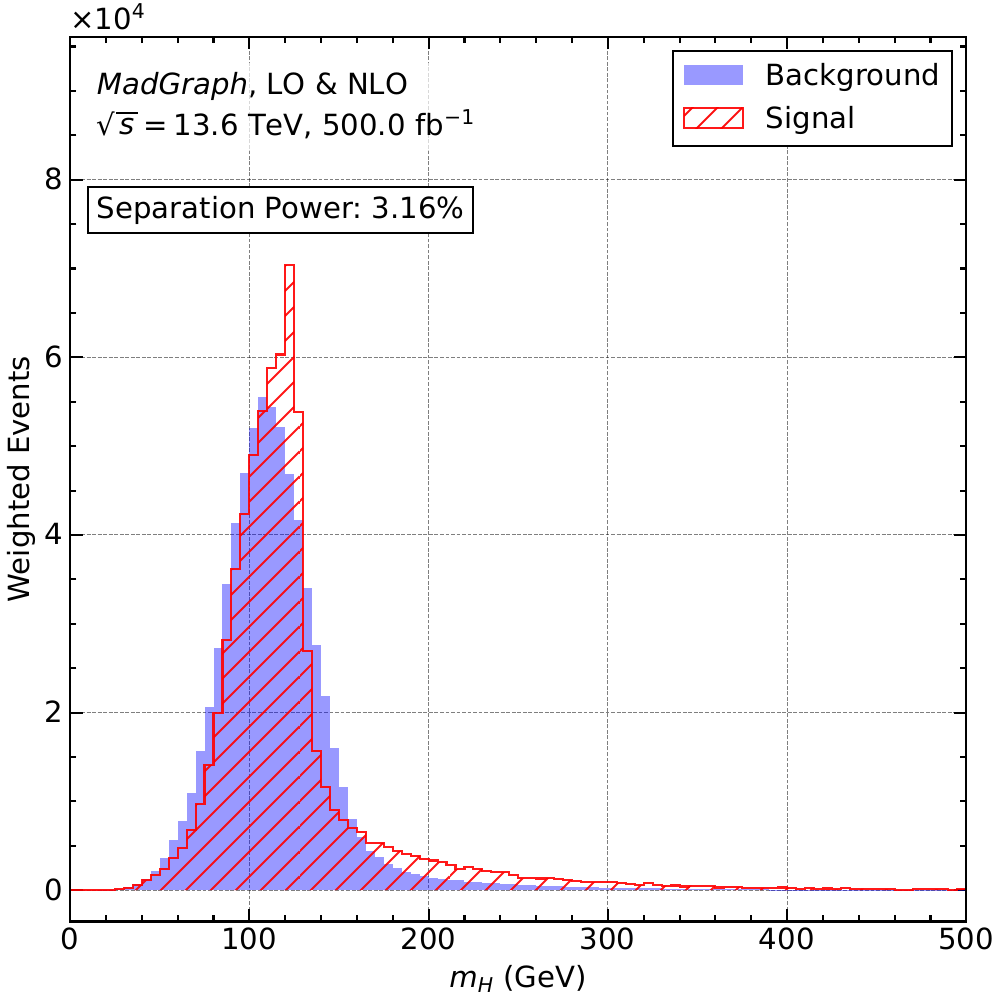}
\qquad
\includegraphics[width=.4\linewidth]{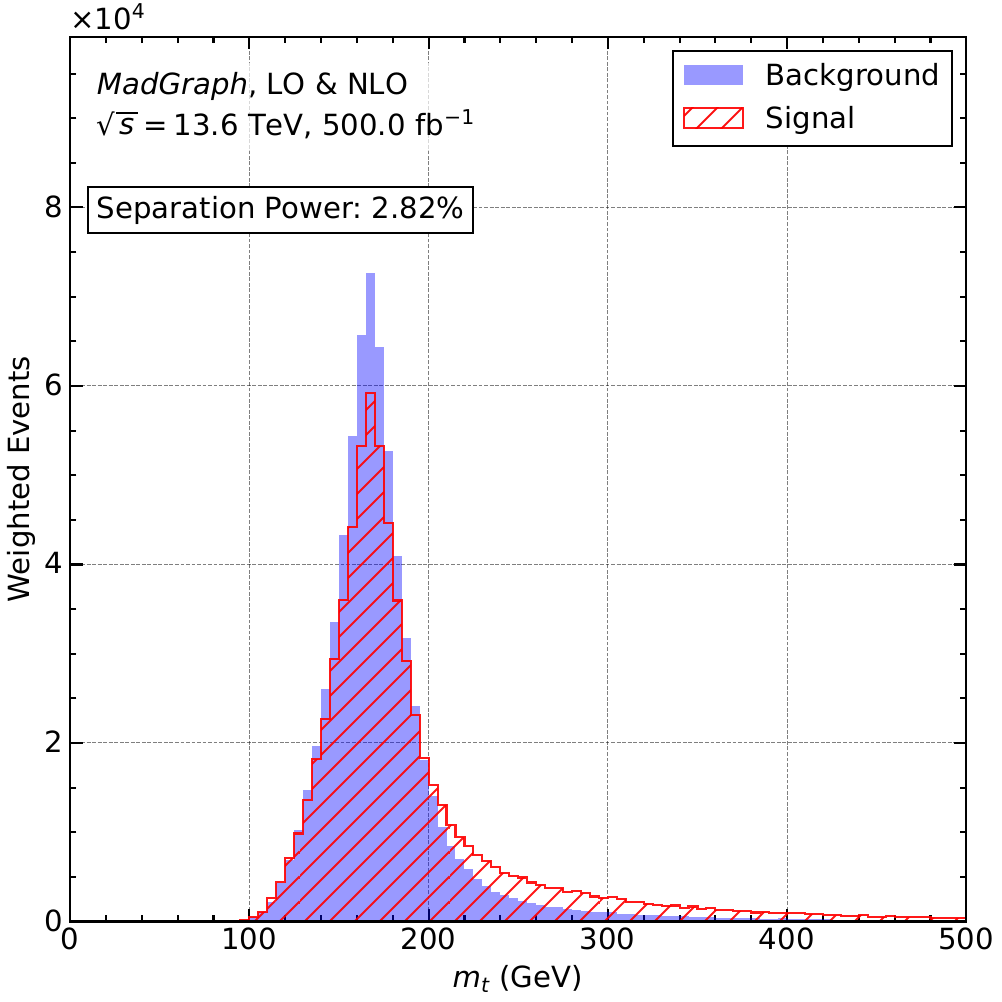}
\caption{The graph on the right illustrates the invariant mass of the reconstructed Higgs boson, while the one on the left illustrates a top-quark for the combination with the smallest $\chi^{2}$.\label{mass_higgs_top}}
\end{figure}

From the three b-jets and the two solutions for the neutrino momentum, six combinations can be formed to reconstruct the top quark and Higgs boson masses. For each combination, a $\chi^{2}$ value was calculated using the formula:

\begin{equation}
    \chi^{2} = (\frac{m_{bb} - m_{H}}{\sigma_{H}})^{2} + (\frac{m_{bl\nu} - m_{t}}{\sigma_{t}})^{2},
\end{equation}
\noindent
where the parameters $m_{H}, m_{t}, \sigma_{H}, \sigma_{t}$ were determined by fitting the distributions for the reconstructed masses corresponding to the correct combination. The reconstructed masses were determined from the combination corresponding to the smallest $\chi^{2}$. The distributions of the invariant mass of the Higgs boson and the top-quark, shown in figure~\ref{mass_higgs_top}, demonstrate the difficulty of separating the signal from the total background using rectangular cuts. The \textit{Separation Power}~\cite{10.48550/arxiv.physics/0703039}, defined in equation~\ref{eq:separation-power}, was used to quantify the difference between the signal and background distributions of an individual variable over its full range. The signal and background histograms were constructed using event weights that account for the process cross-sections, the integrated luminosity, and the sign of the generator weight for each event.

\begin{equation}
\label{eq:separation-power}
S=\frac{1}{2}\sum_{i=1}^{N_{\mathrm{bins}}}\frac{\left(s_i-b_i\right)^2}{s_i+b_i} \,,
\end{equation}
\noindent
where $N_{\mathrm{bins}}$ is the number of histogram bins, and $s_i$ and $b_i$ denote the weighted number of events for signal and background in the $i$-th bin, respectively. Larger values of $S$ indicate greater discriminating power.

\begin{figure}[htbp]
\centering
\includegraphics[width=.4\linewidth]{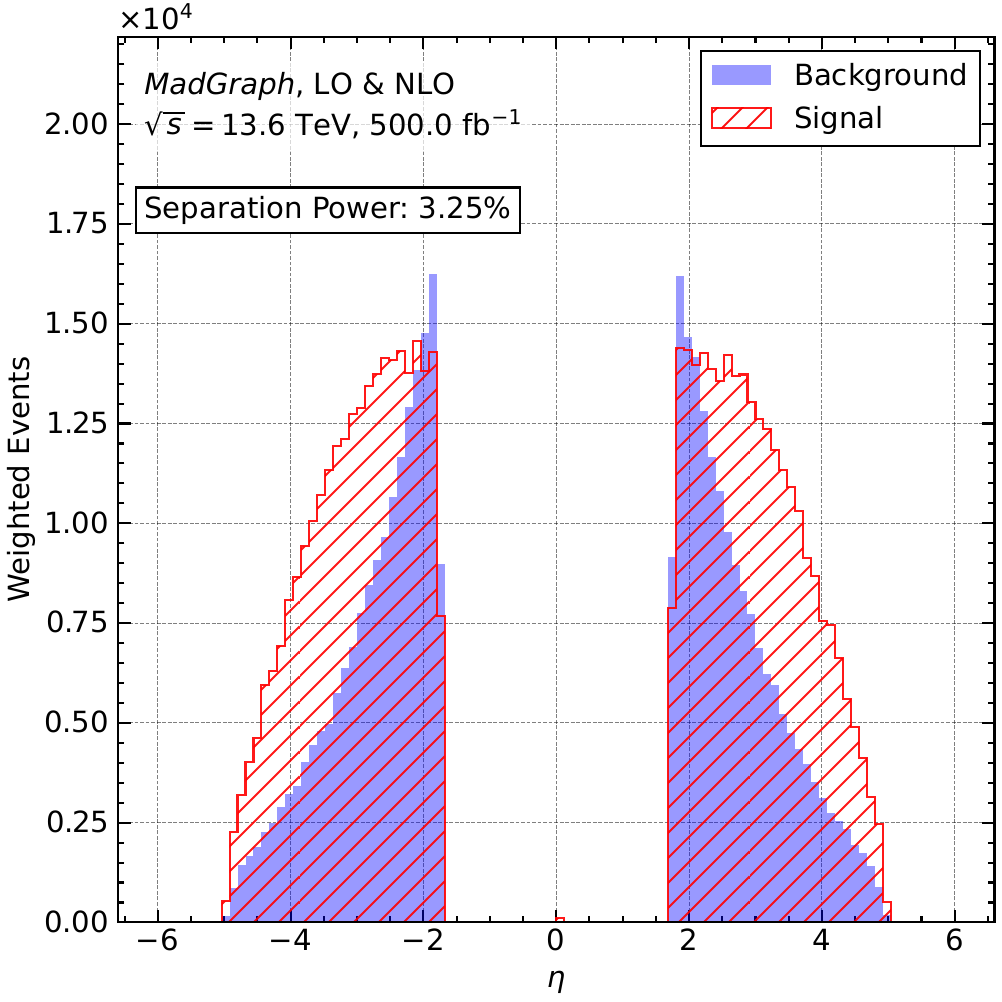}
\qquad
\includegraphics[width=.4\linewidth]{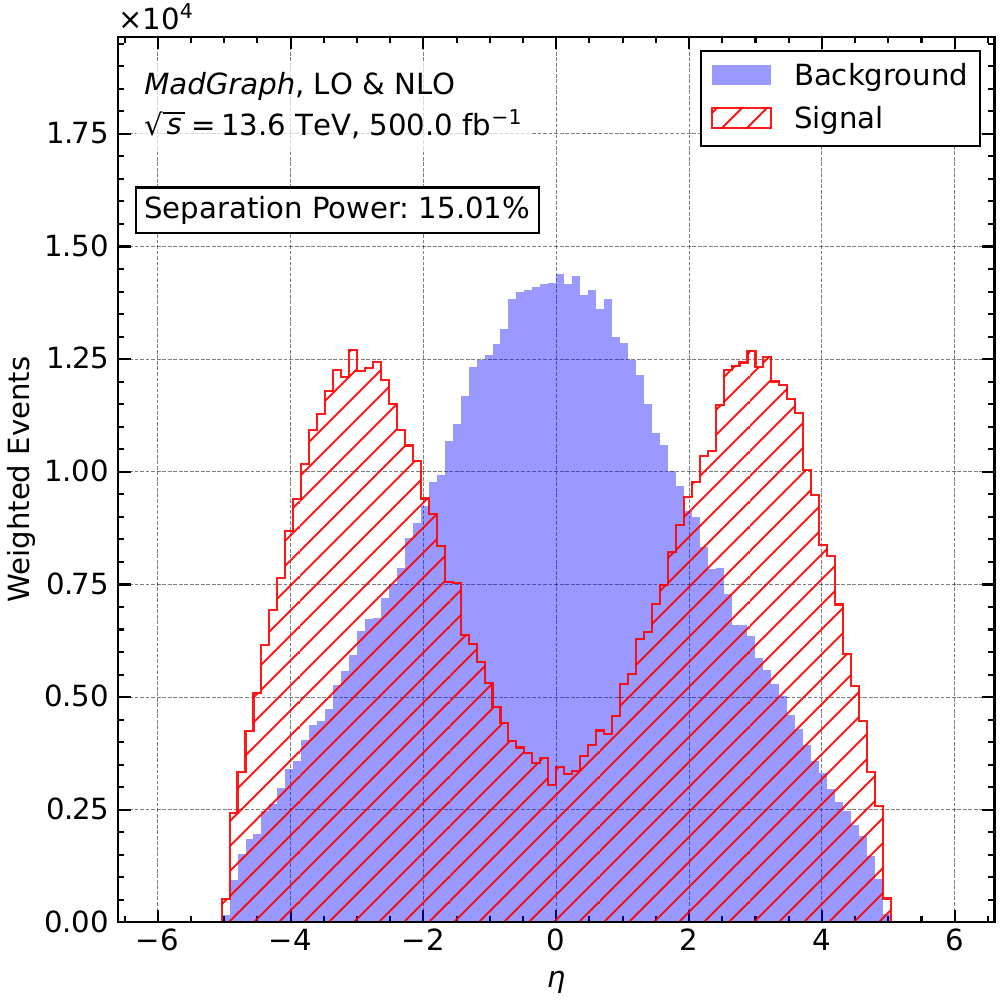}
\caption{Pseudorapidity of the forward jet (left panel) and the tagging jet (right panel).\label{eta_fwd_tag}}
\end{figure}

\begin{figure}[htbp]
\centering
\includegraphics[width=.4\linewidth]{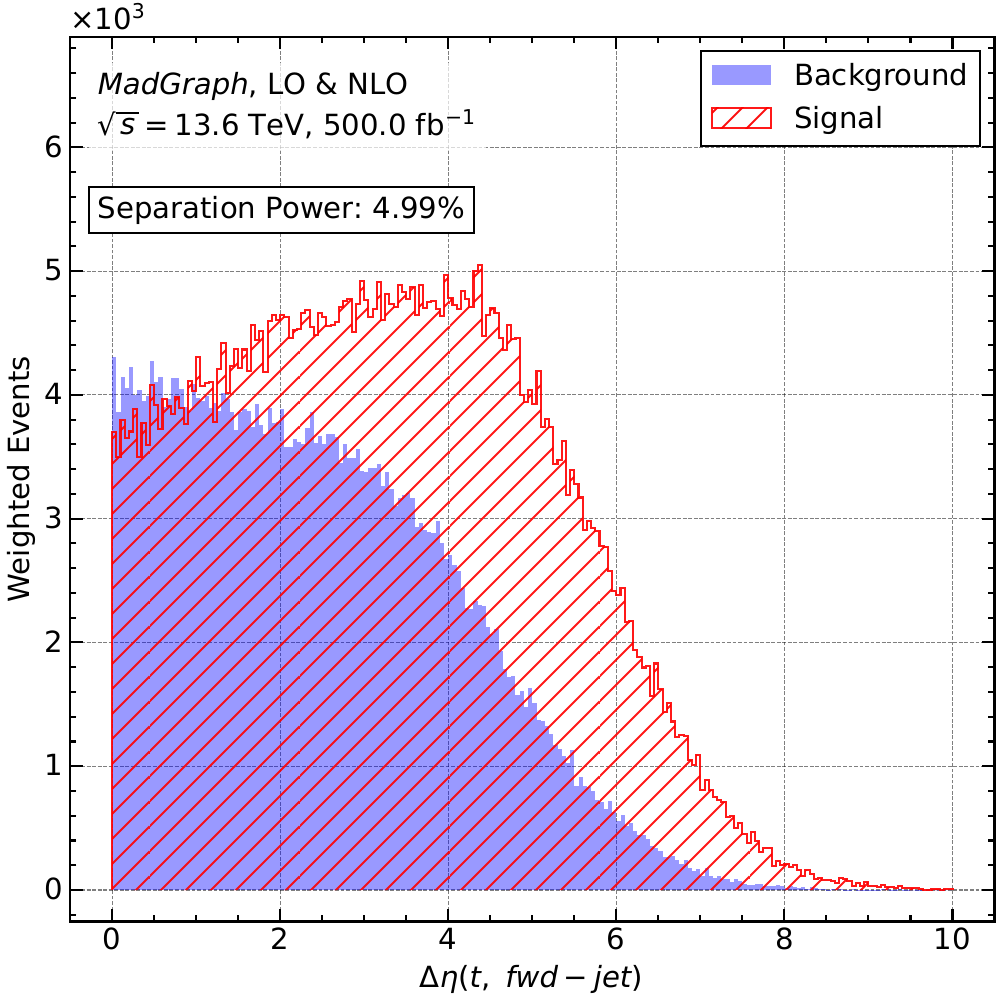}
\qquad
\includegraphics[width=.4\linewidth]{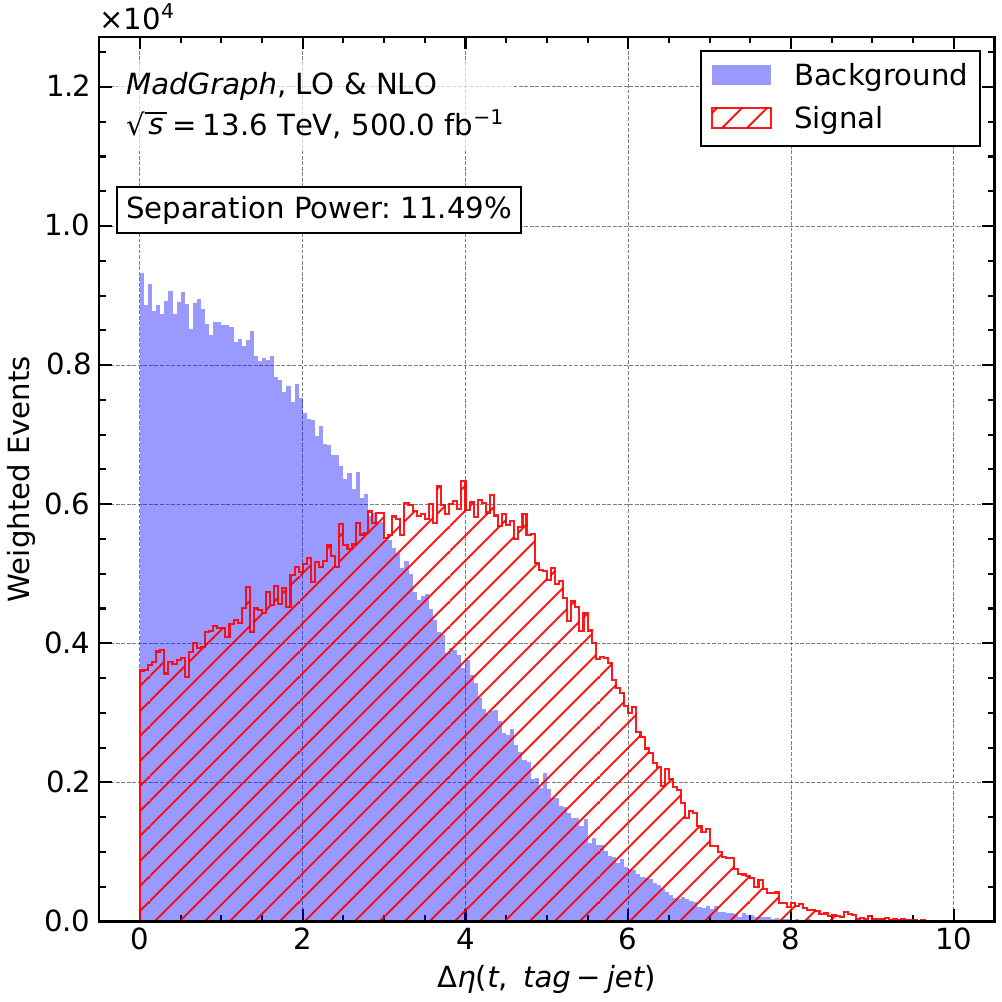}
\caption{Distance in $\eta$ between top-quark and forward jet (left panel), top-quark and tagging jet (right panel).\label{delta_eta_top_fwd_tag}}
\end{figure}

\begin{figure}[htbp]
\centering
\includegraphics[width=.4\linewidth]{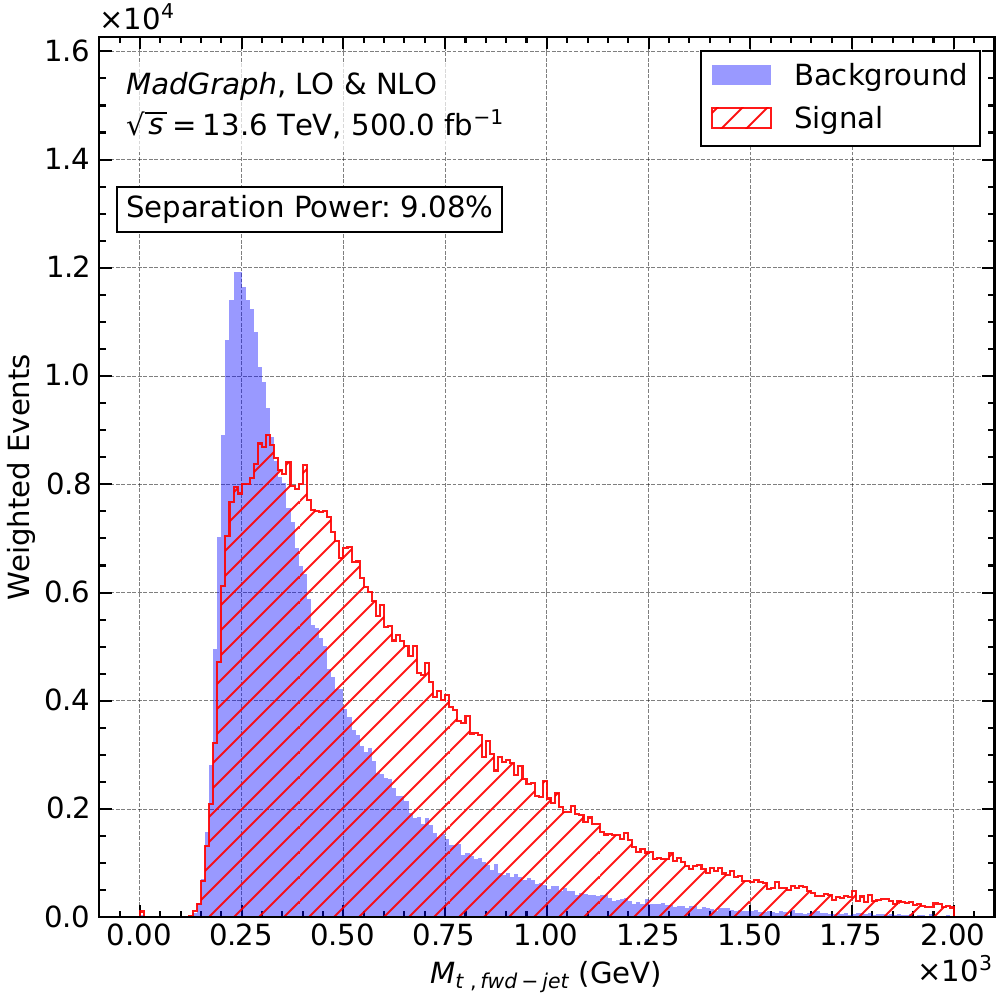}
\qquad
\includegraphics[width=.4\linewidth]{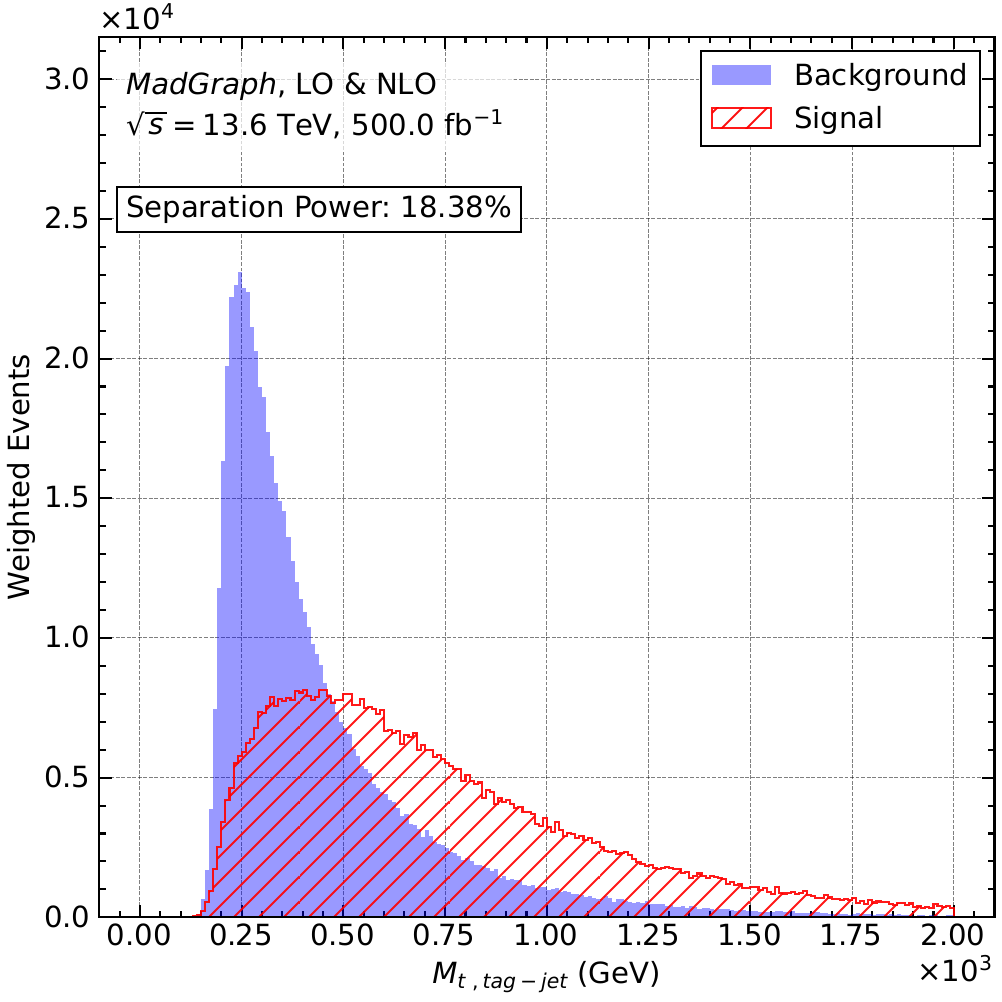}
\caption{Invariant mass of top-quark and forward jet (left panel), top-quark and tagging jet (right panel).\label{mass_top_fwd_tag}}
\end{figure}

In the $pp\rightarrow tH$ process, a "spectator quark" $q'$ is produced (figure~\ref{fig:tH_channels} ), whose momentum is predominantly directed towards regions of high absolute pseudorapidity (typically $|\eta| > 2.0$). The jet originating from this quark can be used to discriminate the signal from the main background process $pp\rightarrow t\bar{t}$, where the forward jet production is not preferred \cite{10.1134/s1063778822020041}. We tested two approaches to identify the jet from the spectator quark: selection of the "forward jet" with $p_{T}^{FW}>30$ GeV/c, $|\eta|>1.74$, and selection of the "tagging jet", i.e. a non-$b$ jet having the largest invariant mass with the $b$-jet from the top-quark decay \cite{10.1007/jhep10(2025)093}. The pseudorapidity distributions of the ”forward” jet selected by the two algorithms are presented in figure~\ref{eta_fwd_tag}. The use of the tagging jet provides a better discrimination between signal and background events, as is evident from the distributions shown in figure ~\ref{delta_eta_top_fwd_tag}-\ref{mass_top_fwd_tag}.

Fifty kinematic variables were constructed and considered for the further training of the neural network that separates signal from background. The full list and description of the kinematic variables are provided in appendix~\ref{app:variables-description}.

\section{Exploratory data analysis}
\label{sec:preprocessing}

Prior to supplying the input features into the machine learning model, several preprocessing steps were performed. The preprocessing stages described below were applied in a specified sequential order to preserve consistency of the data transformation procedure.

\subsection{Correlation analysis}

\begin{figure}[htbp]
\centering
\includegraphics[width=.7\linewidth]{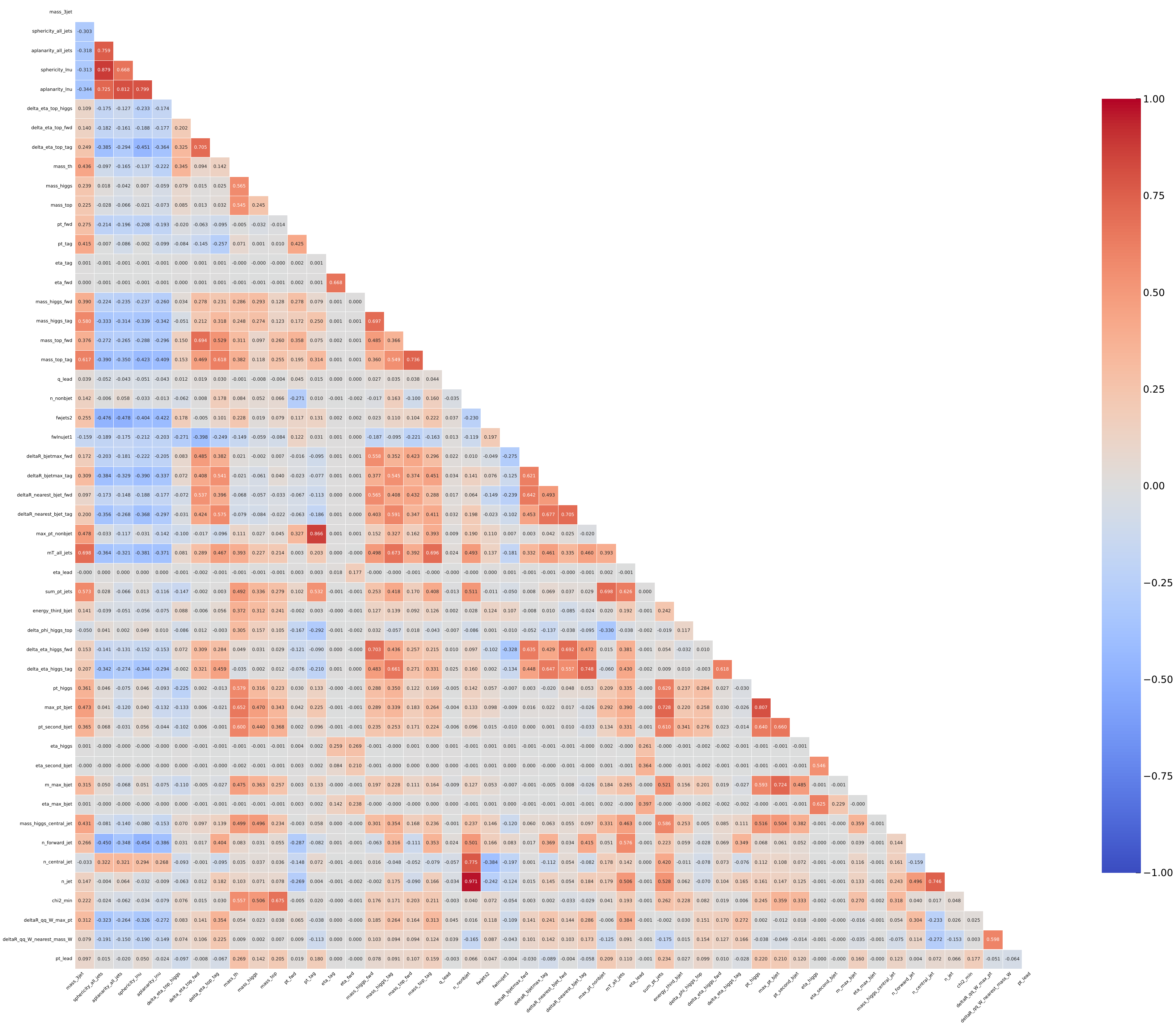}
\caption{Correlation matrix for total events.\label{fig:correlation}}
\end{figure}

First, a feature \textit{Correlation Matrix} (see figure~\ref{fig:correlation}) was constructed, which allowed the identification of five pairs of variables with the Pearson correlation coefficient \cite{10.1098/rspl.1895.0041, 10.1213/ane.0000000000002864} greater than or equal to $\rho \ge 0.8$. Pairs of highly linearly correlated variables are shown in table~\ref{tab:corr_pair}. From each pair of strongly correlated variables only one feature was retained - specifically the one exhibiting the higher \textit{Separation Power} (see equation~\ref{eq:separation-power}).

\begin{table}[htbp]
    \centering
    \begin{tabular}{l|c|c|c}
    \hline
    \textbf{\textnumero} & \textbf{Variable 1} & \textbf{Variable 2} & \textbf{Pearson coefficient} \\
    \hline
    1 & \texttt{n\_nonbjet} & \texttt{n\_jet}$^{*}$ & 0.971 \\
    2 & \texttt{sphericity\_all\_jets} & \texttt{sphericity\_lnu}$^{*}$ & 0.879 \\
    3 & \texttt{pt\_tag} & \texttt{max\_pt\_nonbjet}$^{*}$ & 0.866 \\
    4 & \texttt{aplanarity\_all\_jets}$^{*}$ & \texttt{aplanarity\_lnu} & 0.812 \\
    5 & \texttt{pt\_higgs}$^{*}$ & \texttt{max\_pt\_bjet} & 0.807 \\
    \hline
    \end{tabular}
\caption{Pairs of highly correlated variables. Variables marked with an asterisk ($^{*}$) are excluded from the final input feature set.\label{tab:corr_pair}}
\end{table}

\subsection{Normalization}

To promote proportionate contributions of all input variables during training, feature normalization was implemented. Normalization refers to the linear transformation of a variable such that its values are mapped onto a fixed interval $[0,1]$, while preserving the overall shape of its distribution.

\subsection{Dataset splitting}

The initial dataset was partitioned into training ($70\%$), validation ($15\%$), and testing ($15\%$) subsets. A stratified split was employed based on the process label of each event to preserve the relative proportions of all individual processes across all subsets.

\section{Development and optimization of the neural network}
\label{sec:optimization}

\subsection{Mathematical formulation of the problem}

Given a finite set of processed variables, the objective of this study could be formulated as the reconstruction of an accurate approximation of the likelihood ratio based on simulated labeled events. Although the reconstructed likelihood ratio $\hat{\Lambda}(\mathbf{x})$ could, in principle, be obtained through independent estimation of the signal and background probability density functions, $\hat{p}(S|\mathbf{x})$ and $\hat{p}(B|\mathbf{x})$, respectively, such density estimation becomes statistically intractable in high-dimensional feature spaces due to the \emph{curse of dimensionality}. The number of events required for a reliable estimation of a probability density in a $d$-dimensional space scales exponentially as $N \sim \mathcal{O}(k^d)$, where $k$ denotes the effective binning granularity per dimension. Consequently, direct density reconstruction becomes impractical due to the unrealistically large event statistics required.

Instead of estimating the likelihood ratio directly, it is often more accurate and convenient to express the classification problem in terms of the posterior probability $P(S|\mathbf{x})$ that an event with observed kinematic features $\mathbf{x}$ originates from the signal process. These two quantities are related through Bayes’ theorem as

\begin{equation}
P(S|\mathbf{x})=\frac{\Lambda(\mathbf{x})P(S)}{\Lambda(\mathbf{x})P(S)+P(B)}
\end{equation}
\noindent
where $P(S)$ and $P(B)$ denote the prior probabilities of the signal and background processes, respectively. From this relation it follows that the posterior probability $P(S|\mathbf{x})$ and the likelihood ratio $\Lambda(\mathbf{x})$ represent monotonic transformations of one another. Therefore, both quantities contain equivalent information from the perspective of the classification problem. However, the direct approximation of the posterior probability $P(S|\mathbf{x})$ constitutes a statistically more stable learning task.

To address this problem, a machine learning model with strong nonlinear approximation capabilities was employed, namely a deep neural network \cite{10.1142/12294, Goodfellow-et-al-2016, 10.1038/nature14539}. From a mathematical perspective, a neural network represents a parameterized nonlinear mapping $f(\mathbf{x};\theta)$, where $\theta$ denotes the set of trainable model parameters (weights and biases). Training consists of iteratively adjusting the parameters $\theta$ in order to minimize a loss function defined on the training dataset. For supervised training with the backpropagation algorithm, each event is assigned a binary target label: $f(\mathbf{x}) = 1$ for signal events, while $f(\mathbf{x}) = 0$ is assigned to background events.

With a sigmoid activation function in the output neuron $f(\mathbf{x})\in [0,1]$, the natural choice of loss function is the binary cross-entropy. In the idealized limit, the function minimizing the expected binary cross-entropy risk coincides with the true posterior probability of the signal class, i.e. $f^*(\mathbf{x})=P(S\mid\mathbf{x})$ \cite{10.1162/neco.1991.3.4.461}.

During training, event weights accounting for the process cross-sections and integrated luminosity were applied, followed by equalization of the signal and total background yields. Generator-level event weights were not incorporated, in order to avoid negative weights in the training sample and to prevent the associated risk of the loss function from becoming unbounded from below \cite{10.1088/1748-0221/14/08/P08020}. Under this balanced training, the network response also becomes a monotonic function of the likelihood ratio between the effective signal and background distributions. The trained network therefore defines a scalar field $f(\mathbf{x})$ over the phase space, which was used as a nonlinear discriminant to extract the signal events.

\subsection{Masking}

In some events, individual physical objects or categories may be absent. In such cases, the values of the corresponding kinematic variables are undefined. To preserve a fixed input dimensionality of the neural network, all missing values were encoded with a common placeholder value chosen well outside the physically allowed range of the corresponding variables.

To prevent the network from treating the artificial placeholder value as a numerical feature and thereby learning unphysical correlations induced by it, a dedicated mechanism for masking missing values was implemented. The implementation of this approach requires the network input to be first separated into isolated branches. Each branch is associated with a specific object or category and consists of a small fully connected subnetwork. Its input is a set of kinematic variables describing the corresponding entity. This organization allows different entities to be processed independently and the masking to be applied at the level of the entire input group under consideration. This is important because missing values of variables belonging to a given branch occur simultaneously in the same events.

To conditionally control the contribution of a branch depending on the presence of the corresponding entity, the transformation of its hidden representation was modified according to equation~\eqref{eq:masked_representation}.

\begin{equation}
\label{eq:masked_representation}
H_{masked}=M_{\mathrm{out}}\odot \verb|Dropout|[\phi ( \verb|BatchNormalization|\{W(M_{\mathrm{in}}\odot X)+B\})]
\end{equation}
\noindent
here $X$ denotes the input vector of the branch, $W$ and $B$ are the weight matrix and bias vector of its fully connected layer, respectively, and $\phi$ denotes the activation function. The quantities $M_{\mathrm{in}}$ and $M_{\mathrm{out}}$ are binary masks whose dimensions correspond to the input vector and the hidden representation of the branch, respectively. If at least one value in the input group $X$ is equal to the placeholder value, all elements of both masks are set to zero. If the input group contains physically defined values, all elements of both masks are set to one.

This construction provides two important properties of the specialized branch when missing values are supplied to it:

\begin{enumerate}
    \item Formation of a zero-valued hidden representation of the branch.
    \item Absence of updates to the branch weights and biases due to the corresponding gradients from the loss function being zeroed.
\end{enumerate}

Taken together, these properties prevent kinematic information associated with an absent object from being propagated to the subsequent classification transformation, while such an event does not contribute gradients for updating the parameters of the corresponding branch. When physically defined values are provided, both $M_{\mathrm{in}}$ and $M_{\mathrm{out}}$ consist entirely of ones, and equation~\eqref{eq:masked_representation} reduces to the standard transformation of a fully connected branch. Thus, for present objects, the masking mechanism does not alter the standard computations of the network.

Each branch produces its own hidden representation of the corresponding input group. The resulting representations are concatenated into a single vector. This vector is further augmented with a number of binary presence features equal to the number of masked branches, provided separately from the hidden kinematic representations. Each of these features characterizes the presence of the corresponding entity in the event. A binary presence feature is dynamically constructed for each masked branch: a value of $1$ is assigned if the entire input group has physically defined values, and a value of $0$ if all input variables take the placeholder value. The explicit provision of this information supplies the classifier with additional discriminating information independent of the corresponding kinematic representation.

\begin{figure}[htbp]
\centering
\includegraphics[width=0.3\linewidth]{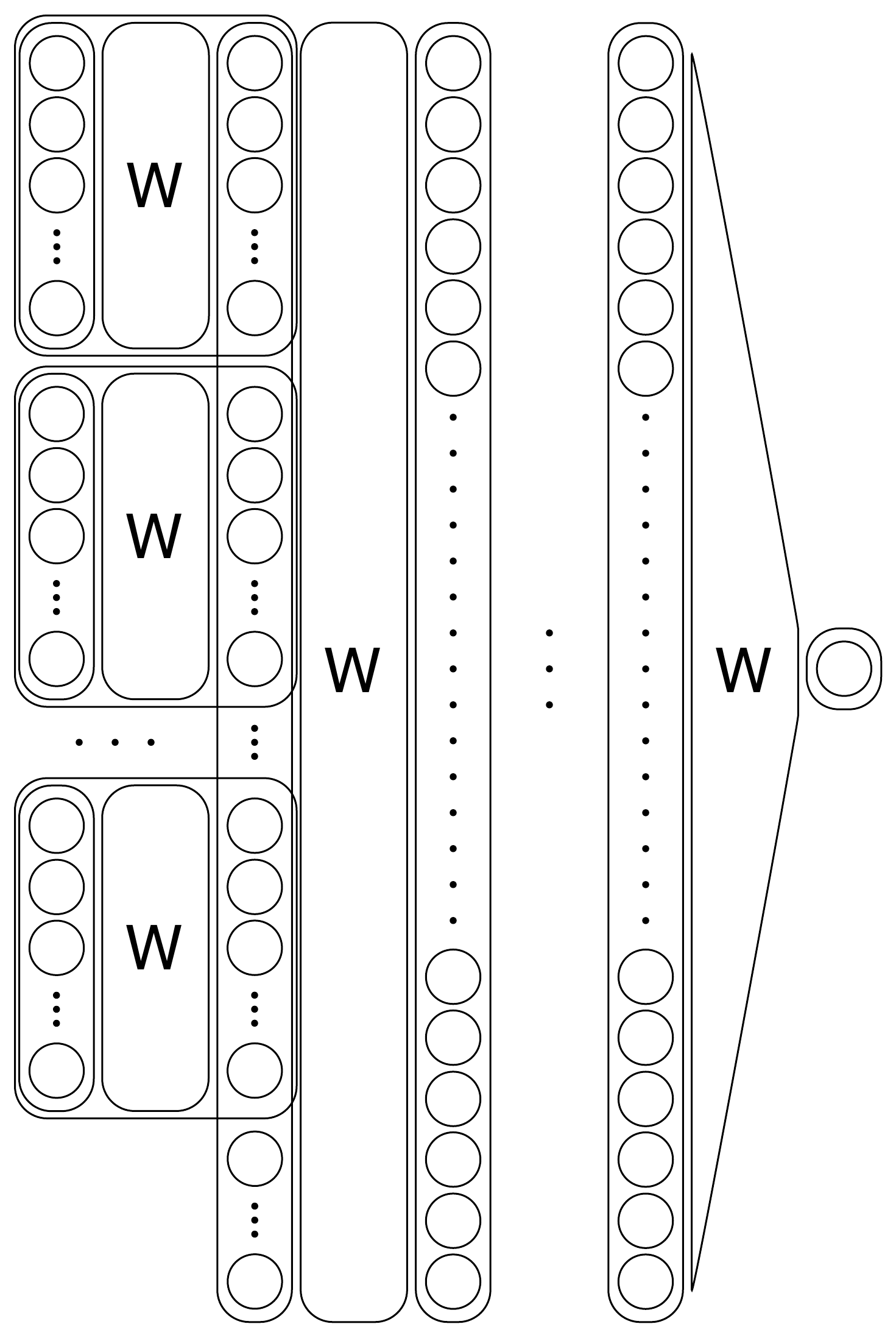}
\caption{General neural network architecture used in the analysis, incorporating the masking mechanism.\label{fig:masked_architecture}}
\end{figure}

The resulting vector is provided as the input to the subsequent common classifier, which is implemented as a fully connected deep neural network. The overall architecture of the proposed neural network is illustrated in figure~\ref{fig:masked_architecture}.

\subsection{Optimization}

According to the universal approximation theorem \cite{10.1007/bf02551274,10.1016/0893-6080(89)90020-8}, a feedforward neural network with at least one hidden layer (perceptron) and a sufficient number of neurons is capable of approximating an arbitrary continuous function on a compact set with a predetermined accuracy. From this perspective, one may assume that almost any network architecture, provided that the hidden layers contain a sufficiently large number of neurons, can achieve the required expressive power.

However, in practice the effectiveness of such approximation is limited by imperfections of the training process, such as the possibility of getting convergence to local minima of the loss function, the risk of overfitting, statistical noise in the data, and limitations of the regularization algorithm. Consequently, training an arbitrary architecture does not guarantee convergence to the global minimum of the loss function and therefore does not ensure an ideal approximation of the objective function as predicted by theory.

Due to the analytically unpredictable influence of the architecture on the geometry of the loss landscape, it is not possible to determine in advance the optimal combination of architectural and training parameters that provides the best approximation quality for the target dependence. To identify the most suitable configuration, an automated hyperparameter optimization procedure \cite{10.1016/j.neucom.2020.07.061, 10.1002/widm.1484} was performed using the \textit{Optuna} library \cite{10.1145/3292500.3330701}. The optimization involves a systematic exploration of all possible predefined combinations of hyperparameters within specified ranges.


An additional step aimed at reducing the computational cost of the optimization procedure was the use of a subsample of events selected from the full dataset that passed the preliminary event selection. The reliability of this approach has been empirically demonstrated in \cite{10.1016/j.heliyon.2024.e26586}. After the initial partitioning (see section~\ref{sec:preprocessing}), a reduced dataset was independently selected from each subset by retaining a fixed number of events for each process. The composition of the resulting training, validation, and testing datasets, containing the same number of events for each process, is summarized in table~\ref{tab:subsample}. Additionally, the $33$ variables containing the dominant discriminating information were selected using the permutation importance method \cite{10.1023/a:1010933404324} for the hyperparameter optimization procedure.

\begin{table}[ht]
    \centering
    \begin{tabular}{l|>{\centering\arraybackslash}p{2.7cm}}
    \hline
    \textbf{Process} & \textbf{Number of events} \\
    \hline
    $tHbq$ & 50,000 \\
    $t\bar{t}$ & 20,000 \\
    $t\bar{t}b\bar{b}$ & 10,000 \\
    $t\bar{t}H$ & 6,000 \\
    $tZbq$ & 6,000 \\
    $t\bar{t}Z$ & 6,000 \\
    $t\bar{t}W$ & 3,000 \\
    \hline
    \end{tabular}
\caption{Substatistics of processes used for the hyperparameter optimization process. The size of the subsample for each process was chosen such that the actual number of signal events was approximately equal to the total number of background events.\label{tab:subsample}}
\end{table}

Since the maximum achievable signal significance depends on the selection threshold, its evaluation relies on only a fraction of the available event statistics and is therefore sensitive to statistical fluctuations. To objectively evaluate the performance of each candidate architecture and guide the optimization toward the highest achievable accuracy, the area under the receiver operating characteristic curve (AUC) \cite{10.1016/s0031-3203(96)00142-2} was chosen as the target metric. By exploiting the full available signal and background samples, AUC provides a more stable basis for comparing a large number of network architectures.

\begin{figure}[htbp]
\centering
\includegraphics[width=0.5\linewidth]{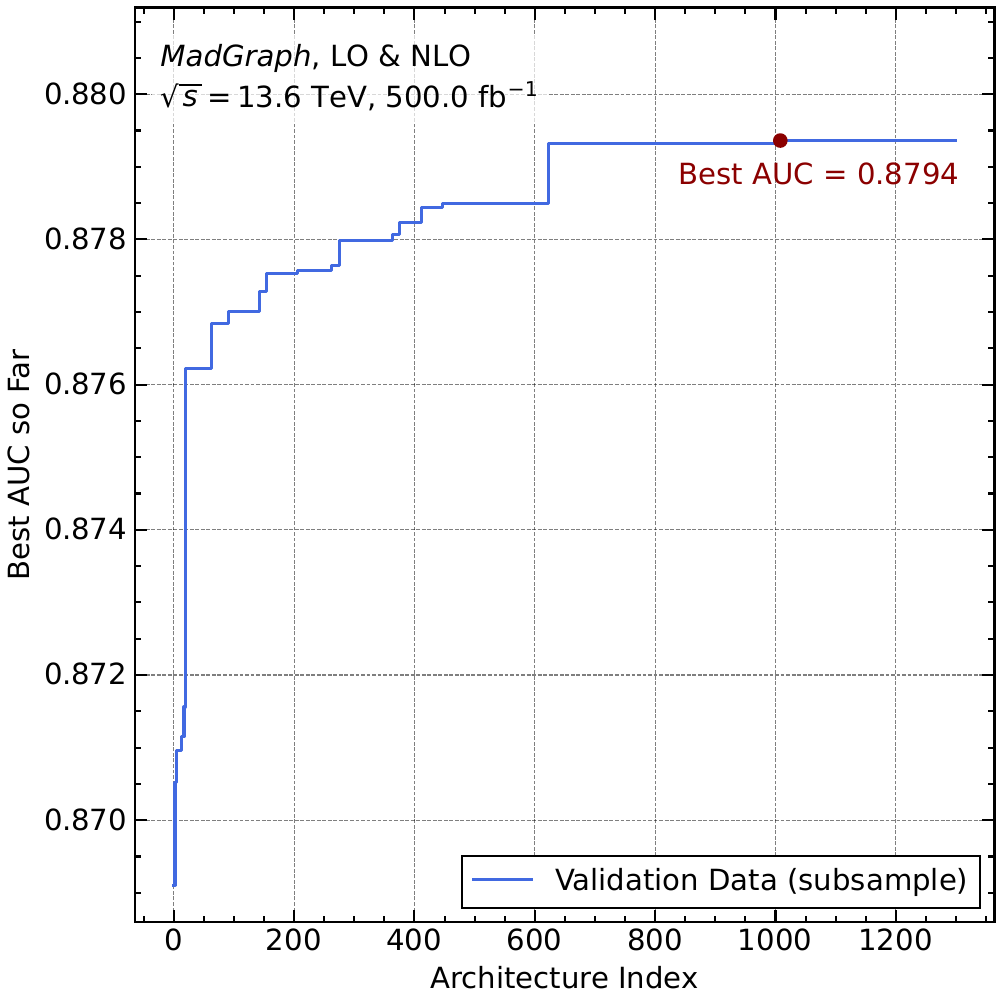}
\caption{Stepwise convergence of the neural network hyperparameter optimization, showing the best $ROC-AUC$ score achieved as a function of the architecture index.\label{fig:optimization_convergence}}
\end{figure}

A Bayesian exploration \cite{frazier2018tutorialbayesianoptimization, 10.48550/ARXIV.1801.01596, 10.11989/JEST.1674-862X.80904120} of all possible combinations was employed in this study — the Tree-structured Parzen Estimator (TPE) algorithm \cite{10.48550/arxiv.2304.11127}. The convergence dynamics of the optimization process is shown in figure~\ref{fig:optimization_convergence}.

Among all $1300$ configurations tested, the best value of the target metric was achieved at the 1009th architecture. Following this point, the algorithm continued exploring the local region of the parameter grid for an additional $291$ iterations. However, no further improvement in the classification performance was observed, indicating that the optimization process had reached a saturation stage. Any potential further improvements in classification accuracy would likely be negligible compared to the accuracy already achieved, given the additional computational resources required. Therefore, it was concluded that terminating the optimization procedure at this stage was justified.

\section{Final model evaluation}
\label{sec:selection}

The architecture and training parameters of the best-performing neural network identified by the hyperparameter optimization are summarized in table~\ref{tab:NN_structure}. The final training of this architecture was performed using the full available event statistics and the complete set of $41$ input kinematic variables. The input variables are described in table~\ref{tab:variables} in the same order as they are provided to the neural network. The same event weighting scheme as described in section~\ref{sec:generation} for the calculation of the separation power metric~\ref{eq:separation-power} was used to construct the evaluation plots.

\begin{table}[htbp]
\centering

\begin{tabularx}{\linewidth}{l|c|c|c|c}
\hline
\multicolumn{5}{c}{\textbf{Branches}}\\
\hline
\textbf{Name} & \textbf{Input neurons} & \textbf{Activation} & \textbf{Dropout} & \textbf{Hidden neurons}\\
\hline
$\mathrm{always\_filled}$ & $24$ & tanh & $0.4$ & $128$\\
$\mathrm{w\_qq}$ & $2$ & mish & $0.2$ & $6$\\
$\mathrm{forward\_jet}$ & $7$ & mish & $0.7$ & $12$\\
$\mathrm{tagging\_jet}$ & $8$ & swish & $0.2$ & $16$\\
\hline
\multicolumn{5}{c}{\textbf{Common classifier}}\\
\hline
\textbf{Layer} & \textbf{Neurons} & \textbf{BatchNorm.} & \textbf{Activation} & \textbf{Dropout}\\
\hline
Input layer & 165 & No & No & No\\
Fully Connected 1 & 224 & Yes & swish & 0.45\\
Fully Connected 2 & 96 & Yes & mish & 0.2\\
Fully Connected 3 & 192 & Yes & relu & 0.25\\
Output Layer & 1 & No & Sigmoid & No \\
\hline
\multicolumn{5}{c}{\textbf{Training settings}}\\
\end{tabularx}

\begin{tabularx}{\linewidth}{c|X|>{\centering\arraybackslash}X}
\hline
\textbf{\textnumero} & \textbf{Parameter} & \textbf{Value}\\
\hline
1 & Loss function & Binary Crossentropy\\
2 & Batch size & 512 \\
3 & Number of epochs & 100\\
4 & Optimizer & Adam \\
5 & Learning rate & $5\times 10^{-4}$ \\
\hline
\end{tabularx}

\caption{
Optimal neural network architecture and training parameters identified through optimization.\label{tab:NN_structure}}
\end{table}

\begin{figure}[htbp]
\centering
\includegraphics[width=0.4\linewidth]{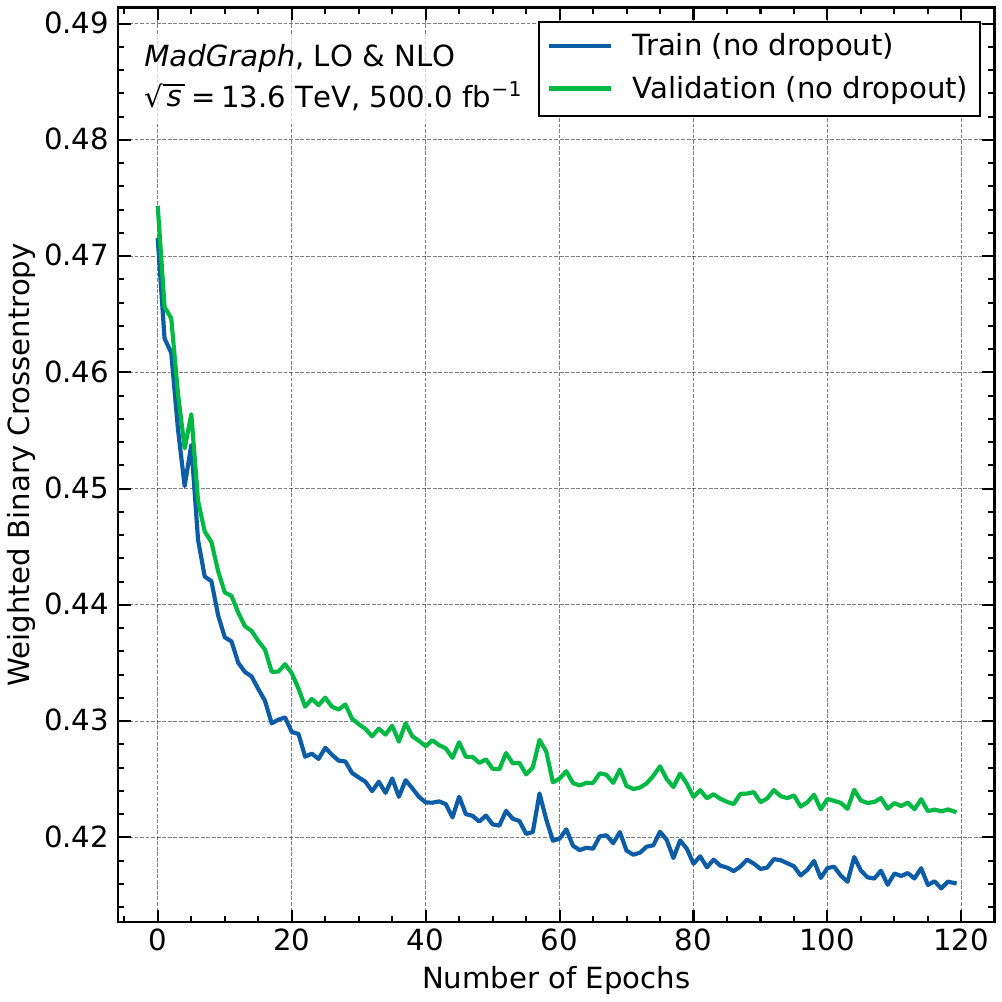}
\caption{The evolution of neural network performance metrics on training and validation events as a function of the number of training epochs.\label{fig:training_history}}
\end{figure}

\begin{figure}[htbp]
\centering
\includegraphics[width=0.5\linewidth]{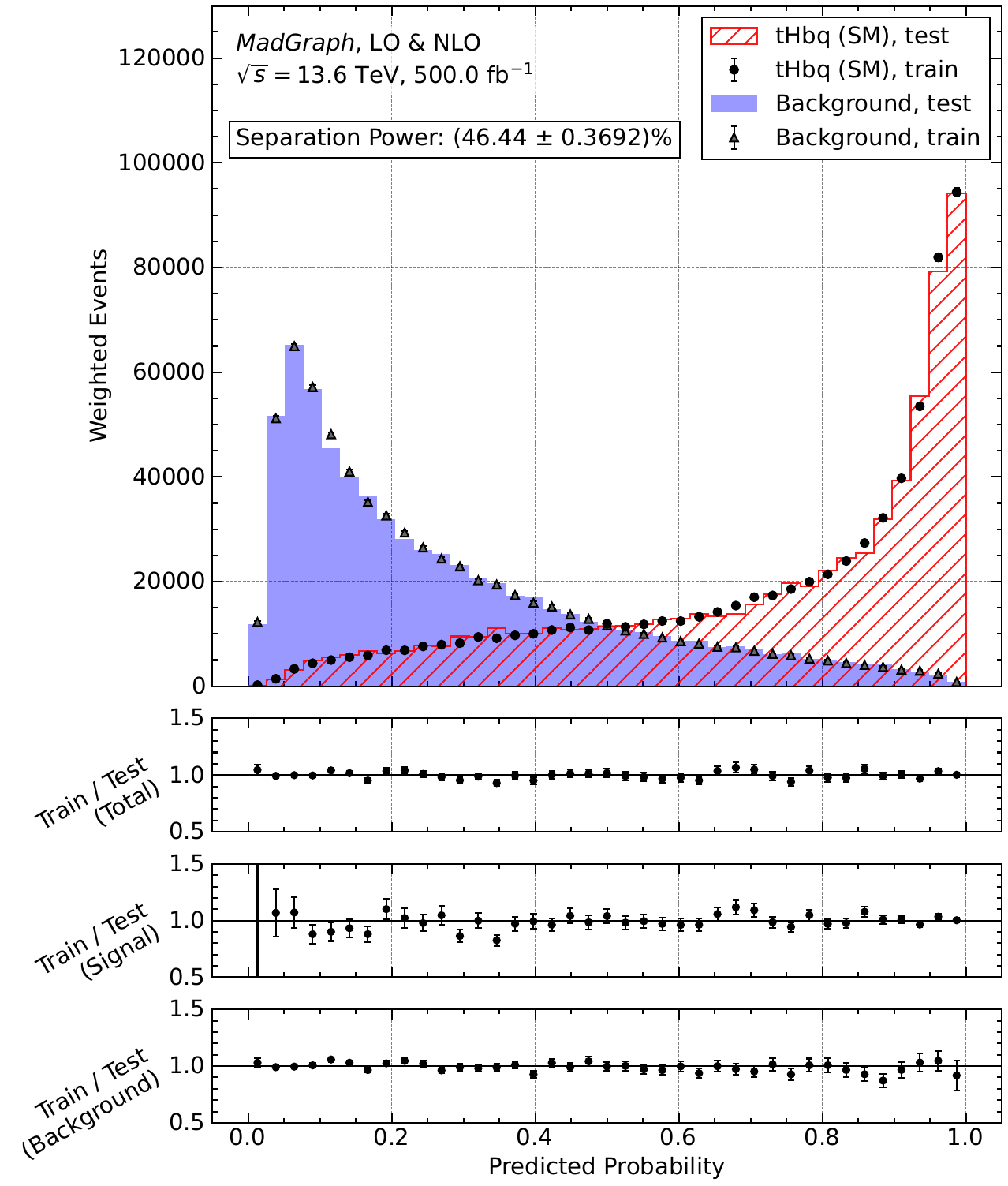}
\caption{Distribution of neural network output predictions. The red dashed histogram represents the likelihood of the network identifying events in which a Higgs boson is produced in association with a single top quark, classifying them as signal events. In contrast, the blue filled histogram corresponds to the probability of background events. The lower panels show the bin-by-bin ratios between the training and testing samples separately for the total event set, the signal process, and the background contribution.\label{fig:nn_prediction}}
\end{figure}

The absence of significant overfitting was assessed based on the lack of substantial divergence between the training and validation curves during training, as shown in figure~\ref{fig:training_history}. The final performance of the neural network was evaluated by visualizing its predictions through a histogram, as shown in figure~\ref{fig:nn_prediction}. The agreement between the network output distributions for the training and testing datasets, reflected in the ratios remaining close to unity over the full range of the output discriminant, further supports the absence of significant overtraining.

\begin{figure}[htbp]
\centering
\includegraphics[width=0.5\linewidth]{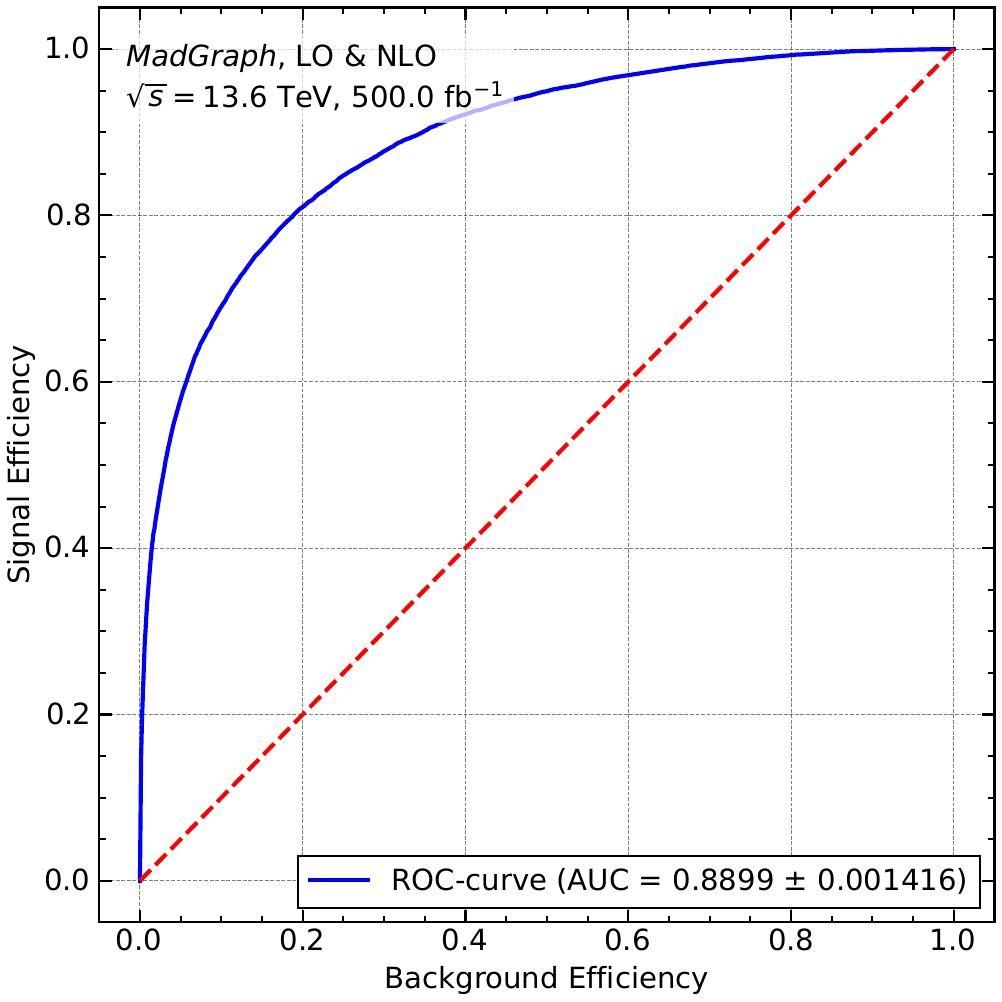}
\caption{ROC-Curve characterizes the overall discriminating power of a binary classifier.\label{fig:nn_roc_curve}}
\end{figure}

To evaluate the overall classification performance of the neural network, the Receiver Operating Characteristic (ROC) curve (see figure~\ref{fig:nn_roc_curve}) was employed. The Area Under the Curve (AUC) metric quantifies the probability that the classifier assigns a higher score to signal than to background events. The achieved classification performance reaches $ROC-AUC = 0.8899 \pm 0.0014$, where the uncertainty due to the limited simulation statistics was estimated using bootstrap method~\cite{10.1214/aos/1176344552}.

\section{Physical sensitivity assessment}
\label{sec:sensitivity}

The statistical signal significance \cite{10.1140/epjc/s10052-011-1554-0} was determined using the Asimov approximation:

\begin{equation}
\label{eq:significance}
Z = \sqrt{2\left[(s+b)\ln\left(1+\frac{s}{b}\right)-s\right]}.
\end{equation}
\noindent
where $s$ represents the expected number of signal events and $b$ denotes the expected number of background events. This metric characterizes the statistical reliability of the observed signal with respect to background fluctuations.

The maximization of the signal significance is accomplished by introducing a threshold $\tau$ in the scalar output space, which defines the decision boundary $f(\mathbf{x})=\tau$. This boundary isolates the signal region $R_s$, where the signal contribution is expected to be statistically most pronounced relative to the background, in accordance with equation~\eqref{eq:significance}.

\begin{figure}[htbp]
\centering
\includegraphics[width=0.5\linewidth]{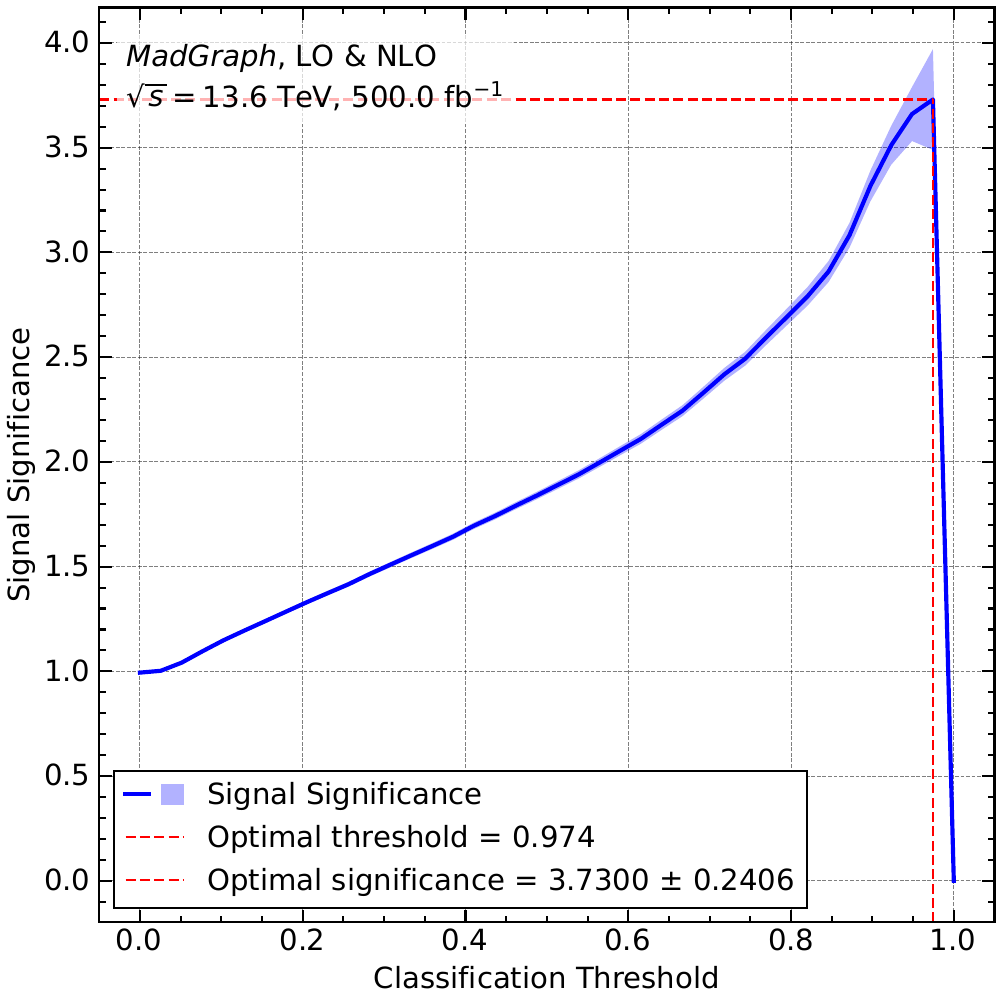}
\caption{Dependence of the expected statistical signal significance on the output threshold of the neural network.\label{fig:significances}}
\end{figure}

To determine the optimal value of the neural network output threshold, the dependence of the signal significance on $\tau$ is studied, as shown in figure~\ref{fig:significances}. For each threshold value, the expected numbers of signal and background events falling within the signal region are evaluated. These values are then used to compute the corresponding signal significance. The associated statistical uncertainty of the signal significance was analytically evaluated using error propagation for indirect measurements.

To avoid signal regions dominated by limited signal statistics, the optimal selection threshold ($\tau$) was required to satisfy two simultaneous requirements, as specified in equation~\eqref{eq:signal_requirements}. If the threshold at which the signal significance reaches its maximum ($\tau_{\mathrm{max}}=\underset{\tau}{\operatorname{arg\,max}}\,Z(\tau)$) did not satisfy all these conditions, the threshold was gradually relaxed until both conditions were fulfilled.

\begin{equation}
\label{eq:signal_requirements}
\frac{\sigma_S}{S}=\frac{\sqrt{\sum_{i\in S}\omega^2_i}}{\sum_{i\in S}\omega_i}\leq 10\%, \qquad S\geq 100.
\end{equation}

Assuming $500~\mathrm{fb}^{-1}$ integrated luminosity, the maximum expected signal significance value of $(3.73 \pm 0.24)\sigma$ satisfying both requirements is achieved at an optimal neural network output threshold value of $\tau=0.974$. This is to be compared with the significance of $0.9937\sigma$ observed at zero threshold, i.e. for all events passing the preselection. The application of this threshold value defines a signal region with a purity of $S/B = 0.128$. The resulting event yields are $113\pm 2$ for the signal and $882\pm 114$ for the background. This selection corresponds to a signal efficiency of $13.75~\%$, while the background rejection reaches $99.87~\%$.

The cut-and-count method selects a single bin in the high-score tail of the classifier predictions distribution. This bin corresponds to a narrow signal-enriched region, resulting in a loss of a fraction of the signal events outside this region. To account for the full available statistics, the complete distribution was divided into $39$ uniformly spaced bins. The expected signal significance was evaluated independently in each bin using the same expression as in equation~\eqref{eq:significance}. The contributions from the individual bins were combined into the overall estimate of the signal significance according to the following prescription:

\begin{equation}  
\label{eq:combined_significance}  
Z =\sqrt{\sum_i Z_i^2}.  
\end{equation}

The application of the shape-fit method resulted in a further increase in the statistical signal significance to $(4.64 \pm 0.20)\sigma$.

\section{Conclusion}
\label{sec:conclusion}

In this work, the expected sensitivity to the production of a Higgs boson in association with a single top quark $tH$ was investigated, focusing on the $H\to b\bar{b}$ decay channel and the semileptonic decay of the top quark $t\rightarrow b l\nu_l$. The analysis was performed using Monte Carlo simulated proton-proton collisions at a centre-of-mass energy of $\sqrt{s}=13.6~\mathrm{TeV}$, assuming an integrated luminosity of $500~\mathrm{fb}^{-1}$.

The initial selection cuts on individual kinematic variables provided effective background suppression. Further signal selection was achieved using a neural network. For the first time, an object-aware masking mechanism was adapted into a deep fully connected neural network architecture for $tH$ search. The input variables were preprocessed to mitigate the influence of anomalous values and improve the suitability of the input variables for network training. The architecture that most accurately approximates the data was determined through hyperparameter optimization.

The best-performing architecture achieved a classification accuracy of $ROC-AUC=0.8899\pm0.0014$. Using the resulting classifier, the expected statistical only significance was estimated to be $(3.73\pm0.24)\sigma$ with the cut-and-count method and $(4.64\pm0.20)\sigma$ with the shape-based fit. The shape-fit method yields a factor of $4.7$ improvement in the signal significance compared to the rectangular-cut-based preselection stage, which achieved a significance of $0.9937\sigma$.

Further improvements to the analysis will involve the implementation of realistic $b$-tagging efficiencies and misidentification rates, accounting for detector response through simulation, and the incorporation of the corresponding systematic uncertainties. It is anticipated that accounting for these additional effects will lead to a significant reduction in the estimated signal significance compared with the current simplified simulation setup. This will provide a more realistic estimate of the sensitivity achievable in an actual LHC analysis.

\appendix
\section{Variables description}
\label{app:variables-description}

The full list of kinematic variables is given in table~\ref{tab:variables}.

\begin{table}[htbp]
\centering
\scriptsize
\makebox[\dimexpr\textwidth-0.5cm\relax][c]{
\begin{tabular}{m{1.3cm}|m{0.5cm}|m{4cm}|m{9.5cm}|m{1cm}}
\hline
\textbf{Category} & \textbf{\textnumero} & \textbf{Encoding} & \textbf{Variable description} & \textbf{SP~(\%)} \\
\hline


\multirow{24}{*}{\hspace{0.5cm}\rotatebox[origin=c]{90}{\texttt{always\_filled}}}
& 1 & \texttt{mass\_3jet}
& Invariant mass of three jets with the largest $p_T$
& 13.22 \\

& 2 & \texttt{sphericity\_all\_jets}
& Measure of uniformity of the distribution of jets in the space
& 14.85 \\

& 3 & \texttt{aplanarity\_lnu}
& Measure of deviation of objects from the common plane
& 14.87 \\

& 4 & \texttt{delta\_eta\_top\_higgs}
& Difference of pseudorapidities from the top quark and Higgs boson
& 6.08 \\

& 5 & \texttt{mass\_th}
& Invariant mass of top quark and Higgs boson
& 4.90 \\

& 6 & \texttt{mass\_higgs}
& Reconstructed mass of Higgs boson
& 3.16 \\

& 7 & \texttt{mass\_top}
& Reconstructed mass of the top quark
& 2.82 \\

& 8 & \texttt{q\_lead}
& Charge of the lepton with the largest $p_{T}$
& 1.55 \\

& 9 & \texttt{n\_nonbjet}
& Number of jets produced by quarks other than a $b$ quark
& 1.34 \\

& 10 & \texttt{fwjets2}
& Second Fox–Wolfram moment composed of the momenta of jets
& 4.64 \\

& 11 & \texttt{fwlnujet1}
& First Fox–Wolfram moment composed of the momenta of objects
& 1.56 \\

& 12 & \texttt{mT\_all\_jets}
& Transverse mass of all jets
& 11.83 \\

& 13 & \texttt{eta\_lead}
& Pseudorapidity of the lepton with the largest $p_{T}$
& 0.03 \\

& 14 & \texttt{sum\_pt\_jets}
& Algebraic sum of transverse momenta of all jets
& 1.27 \\

& 15 & \texttt{delta\_phi\_higgs\_top}
& Difference between the azimuthal angles of the top quark and Higgs boson
& 0.92 \\

& 16 & \texttt{max\_pt\_bjet}
& Transverse momentum of the $b$ jet with the largest $p_{T}$
& 1.09 \\

& 17 & \texttt{pt\_second\_bjet}
& Transverse momentum of the $b$ jet second in its $p_{T}$ value
& 0.52 \\

& 18 & \texttt{eta\_higgs}
& Reconstructed pseudorapidity of Higgs boson
& 0.76 \\

& 19 & \texttt{eta\_second\_bjet}
& Pseudorapidity of the $b$ jet second in its $p_{T}$ value
& 0.80 \\

& 20 & \texttt{m\_max\_bjet}
& Invariant mass of the $b$ jet with the largest $p_{T}$
& 0.17 \\

& 21 & \texttt{eta\_max\_bjet}
& Pseudorapidity of the $b$ jet with the largest $p_{T}$
& 1.01 \\

& 22 & \texttt{n\_central\_jet}
& Number of central light jets
& 11.68 \\

& 23 & \texttt{chi2\_min}
& Quality ($\chi^2$ criterion) of determining the masses of Higgs boson and top quark
& 2.05 \\

& 24 & \texttt{pt\_lead}
& Largest transverse momentum among leptons
& 0.39 \\

\hline


\multirow{2}{*}{\hspace{0.5cm}\rotatebox[origin=c]{90}{\texttt{w\_qq}}}
& 25 & \texttt{deltaR\_qq\_W\_max\_pt}
& Difference between the rapidities of the light jets with the largest $p_{T}$
& 8.87 \\

& 26 & \texttt{deltaR\_qq\_W\_nearest\_mass\_W}
& Difference between the rapidities of the light jets whose masses are closest to the mass of the W boson
& 4.68 \\

\hline


\multirow{7}{*}{\hspace{0.5cm}\rotatebox[origin=c]{90}{\texttt{forward\_jet}}}
& 27 & \texttt{delta\_eta\_higgs\_fwd}
& Difference between the pseudorapidities of Higgs boson and forward jet
& 0.96 \\

& 28 & \texttt{delta\_eta\_top\_fwd}
& Difference of pseudorapidities of the top quark forward jet
& 4.99 \\

& 29 & \texttt{pt\_fwd}
& Transverse momentum of the forward jet
& 3.87 \\

& 30 & \texttt{eta\_fwd}
& Forward jet pseudorapidity
& 3.25 \\

& 31 & \texttt{mass\_top\_fwd}
& Invariant mass of the top quark and forward jet
& 9.08 \\

& 32 & \texttt{deltaR\_bjetmax\_fwd}
& Difference between the pseudorapidities of the forward jet and the $b$ jet with the largest $p_{T}$
& 2.82 \\

& 33 & \texttt{deltaR\_nearest\_bjet\_fwd}
& Difference between the rapidities of the forward jet and the $b$ jet closest to it
& 2.49 \\

\hline

\multirow{8}{*}{\hspace{0.5cm}\rotatebox[origin=c]{90}{\texttt{tagging\_jet}}}
& 34 & \texttt{delta\_eta\_higgs\_tag}
& Difference between the pseudorapidities of Higgs boson and tagging jet
& 4.05 \\

& 35 & \texttt{pt\_tag}
& Transverse momentum of the tagging jet &
1.87 \\

& 36 & \texttt{eta\_tag} &
Tagging jet pseudorapidity
& 15.01 \\

& 37 & \texttt{delta\_eta\_top\_tag}
& Difference of pseudorapidities of the top quark tagging jet
& 11.49 \\

& 38 & \texttt{mass\_higgs\_tag}
& Invariant mass of Higgs boson and tagging jet
& 14.71 \\

& 39 & \texttt{mass\_top\_tag}
& Invariant mass of the top quark and tagging jet
& 18.38 \\

& 40 & \texttt{deltaR\_bjetmax\_tag}
& Difference between the pseudorapidities of the tagging jet and the $b$ jet with the largest $p_{T}$
& 9.63 \\

& 41 & \texttt{deltaR\_nearest\_bjet\_tag}
& Difference between the rapidities of the tagging jet and the $b$ jet closest to it
& 10.96 \\

\hline


& 42 & \texttt{n\_jet}$^{*}$
& Number of jets
& 0.48 \\

& 43 & \texttt{sphericity\_lnu}$^{*}$
& Measure of uniformity of the distribution of objects in the space
& 14.44 \\

& 44 & \texttt{max\_pt\_nonbjet}$^{*}$
& Largest transverse momentum among light jets
& 1.39 \\

& 45 & \texttt{aplanarity\_all\_jets}$^{*}$
& Measure of deviation of the jets from the common plane
& 14.86 \\

& 46 & \texttt{pt\_higgs}$^{*}$
& Reconstructed transverse momentum of Higgs boson
& 0.98 \\

& 47 & \texttt{mass\_higgs\_central\_jet}$^\dagger$
& Invariant mass of Higgs boson and central light jet
& 6.45 \\

& 48 & \texttt{n\_forward\_jet}$^\dagger$
& Number of forward light jets
& 8.59 \\

& 49 & \texttt{mass\_higgs\_fwd}$^\dagger$
& Invariant mass of Higgs boson and forward jet
& 5.18 \\

& 50 & \texttt{energy\_third\_bjet}$^\dagger$
& Energy of the b jet third in its transverse momentum
& 1.40 \\

\hline
\end{tabular}
}
\caption{Variables description. The variable marked with an asterisk ($^{*}$) are excluded due to strong correlations, while variables marked with a dagger ($^\dagger$) are removed based on their negative contribution, as determined by the permutation importance method.\label{tab:variables}}
\end{table}

\acknowledgments

The authors express sincere gratitude to Aleksey Bondyakov for his valuable technical support in configuring the interaction between the local workstation and the remote storage infrastructure of the Laboratory of Information Technologies. This contribution was essential to improve the data management workflow.

Valuable assistance was also provided by Viktoriia Kiseeva through discussions of modern approaches to masking missing values and by providing a demonstrative implementation of this approach, which served as the basis for its subsequent adaptation and application in the present analysis.


\bibliographystyle{biblio/JHEP.bst}
\bibliography{biblio/biblio.bib}

@article{10.1103/physrevlett.19.1264,
  title = {A Model of Leptons},
  volume = {19},
  issn = {0031-9007},
  url = {https://dx.doi.org/10.1103/PhysRevLett.19.1264},
  doi = {10.1103/physrevlett.19.1264},
  issue = {21},
  journal = {Physical Review Letters},
  publisher = {American Physical Society (APS)},
  author = {Weinberg,  Steven},
  year = {1967},
  month = nov,
  pages = {1264–1266}
}

@inbook{10.1142/9789812795915_0034,
  title = {Weak and electromagnetic interactions},
  isbn = {9789812795915},
  issn = {1793-1207},
  url = {http://dx.doi.org/10.1142/9789812795915_0034},
  doi = {10.1142/9789812795915_0034},
  booktitle = {Selected Papers of Abdus Salam},
  publisher = {WORLD SCIENTIFIC},
  author = {Salam,  Abdus},
  year = {1994},
  month = may,
  pages = {244–254}
}

@article{10.1103/PhysRevLett.13.321,
  title = {{Broken Symmetry and the Mass of Gauge Vector Mesons}},
  author = {Englert, F. and Brout, R.},
  journal = {Physical Review Letters},
  volume = {13},
  issue = {9},
  ISSN = {0031-9007},
  pages = {321--323},
  numpages = {0},
  year = {1964},
  month = {August},
  publisher = {American Physical Society (APS)},
  doi = {10.1103/PhysRevLett.13.321},
  url = {http://dx.doi.org/10.1103/PhysRevLett.13.321}
}

@article{10.1103/PhysRevLett.13.508,
  title = {{Broken Symmetries and the Masses of Gauge Bosons}},
  author = {Higgs, Peter W.},
  journal = {Physical Review Letters},
  volume = {13},
  ISSN = {0031-9007},
  issue = {16},
  pages = {508--509},
  numpages = {0},
  year = {1964},
  month = {October},
  publisher = {American Physical Society (APS)},
  doi = {10.1103/PhysRevLett.13.508},
  url = {http://dx.doi.org/10.1103/PhysRevLett.13.508}
}

@article{10.1016/j.physletb.2012.08.020,
author={{ATLAS Collaboration and others}},
title = {Observation of a new particle in the search for the {Standard Model Higgs} boson with the {ATLAS} detector at the {LHC}},
publisher = {Elsevier BV},
journal = {Physics Letters B},
volume = {716},
number = {1},
pages = {1-29},
year = {2012},
month = {September},
issn = {0370-2693},
doi = {10.1016/j.physletb.2012.08.020},
url = {https://www.sciencedirect.com/science/article/pii/S037026931200857X}
}

@article{10.1016/j.physletb.2012.08.021,
author={{CMS Collaboration} and others},
title = {Observation of a new boson at a mass of 125 {GeV} with the {CMS} experiment at the {LHC}},
publisher = {Elsevier BV},
journal = {Physics Letters B},
volume = {716},
number = {1},
pages = {30-61},
year = {2012},
month = {September},
issn = {0370-2693},
doi = {10.1016/j.physletb.2012.08.021},
url = {https://www.sciencedirect.com/science/article/pii/S0370269312008581}
}

@article{10.1103/PhysRevLett.76.4468,
  title = {Determining the CP Nature of a Neutral Higgs Boson at the CERN Large Hadron Collider},
  author = {Gunion, John F. and He, Xiao-Gang},
  journal = {Physical Review Letters},
  volume = {76},
  ISSN = {1079-7114},
  issue = {24},
  pages = {4468--4471},
  numpages = {0},
  year = {1996},
  month = {June},
  publisher = {American Physical Society (APS)},
  doi = {10.1103/PhysRevLett.76.4468},
  url = {http://dx.doi.org/10.1103/PhysRevLett.76.4468}
}

@article{10.1142/S0217751X15501560,
author = {He, Xiao-Gang and Li, Guan-Nan and Zheng, Ya-Juan},
title = {{Probing Higgs boson CP properties with $t\bar{t}H$ at the LHC and the 100 TeV pp collider}},
journal = {International Journal of Modern Physics A},
volume = {30},
number = {25},
pages = {1550156},
year = {2015},
doi = {10.1142/S0217751X15501560},
URL = {https://doi.org/10.1142/S0217751X15501560},
eprint = {https://doi.org/10.1142/S0217751X15501560}
}

@article{10.1088/0954-3899/41/11/113001,
  title = {{Precision measurements of Higgs couplings: implications for new physics scales}},
  volume = {41},
  ISSN = {1361-6471},
  url = {http://dx.doi.org/10.1088/0954-3899/41/11/113001},
  DOI = {10.1088/0954-3899/41/11/113001},
  number = {11},
  journal = {Journal of Physics G: Nuclear and Particle Physics},
  publisher = {IOP Publishing},
  author = {Englert,  C and Freitas,  A and M\"{u}hlleitner,  M M and Plehn,  T and Rauch,  M and Spira,  M and Walz,  K},
  year = {2014},
  month = {September},
  pages = {113001}
}

@article{10.1103/PhysRevD.63.014018,
  title = {Single top quark production as a window to physics beyond the standard model},
  author = {Tait, Tim M. P. and Yuan, C.-P.},
  ISSN = {1089-4918},
  journal = {Physical Review D},
  volume = {63},
  issue = {1},
  pages = {014018},
  numpages = {15},
  year = {2000},
  month = {December},
  publisher = {American Physical Society (APS)},
  doi = {10.1103/PhysRevD.63.014018},
  url = {http://dx.doi.org/10.1103/PhysRevD.63.014018}
}

@article{10.1103/PhysRevD.81.034020,
  title = {Single top and Higgs associated production at the LHC},
  author = {Barger, Vernon and McCaskey, Mathew and Shaughnessy, Gabe},
  journal = {Physical Review D},
  volume = {81},
  ISSN = {1550-2368},
  issue = {3},
  pages = {034020},
  numpages = {9},
  year = {2010},
  month = {February},
  publisher = {American Physical Society (APS)},
  doi = {10.1103/PhysRevD.81.034020},
  url = {http://dx.doi.org/10.1103/PhysRevD.81.034020}
}

@article{10.1103/physrevd.64.094023,
title = {{Associated production of the Higgs boson and a single top quark at hadron colliders}},
author = {Maltoni, F. and Paul, K. and Stelzer, T. and Willenbrock, S.},
journal = {Physical Review D},
issn = {1089-4918},
volume = {64},
number = {9},
pages = {094023},
numpages = {12},
year = {2001},
month = Oct,
publisher = {American Physical Society},
doi = {10.1103/physrevd.64.094023},
url = {https://dx.doi.org/10.1103/physrevd.64.094023},
}

@article{10.1140/epjc/s10052-015-3475-9,
  title = {{Higgs production in association with a single top quark at the LHC}},
  volume = {75},
  ISSN = {1434-6052},
  url = {http://dx.doi.org/10.1140/epjc/s10052-015-3475-9},
  DOI = {10.1140/epjc/s10052-015-3475-9},
  number = {6},
  journal = {The European Physical Journal C},
  publisher = {Springer Science and Business Media LLC},
  author = {Demartin,  Federico and Maltoni,  Fabio and Mawatari,  Kentarou and Zaro,  Marco},
  year = {2015},
  month = Jun
}

@article{10.1051/epjconf/201920104003,
  title = {{Higgs boson production in association with a single top quark at the LHC}},
  volume = {201},
  ISSN = {2100-014X},
  url = {http://dx.doi.org/10.1051/epjconf/201920104003},
  DOI = {10.1051/epjconf/201920104003},
  journal = {EPJ Web of Conferences},
  publisher = {EDP Sciences},
  author = {Koval,  Oksana A. and Boyko,  Igor R. and Huseynov,  Nazim},
  editor = {Bobrikov,  I. and Chudoba,  V. and Derenovskaya,  O. and Friesen,  A. and Verkheev,  A.},
  year = {2019},
  pages = {04003}
}

@misc{CERN_HiggsXSBR,
author = {{LHC Higgs Working Group}},
title = {Higgs cross sections and decay branching ratios},
howpublished = {\url{https://twiki.cern.ch/twiki/bin/view/LHCPhysics/HiggsXSBR}},
note = {[Online; accessed 09-October-2023]}
}

@article{10.1134/s1063778822020041,
author = {I.R. Boyko and N.A. Huseynov and O.A. Koval},
title = {{Monte Carlo Study of Associated Higgs Boson Production with a Single Top Quark}},
volume = {85},
issn = {1562-692X},
url = {http://dx.doi.org/10.1134/S1063778822020041},
doi = {10.1134/s1063778822020041},
number = {2},
journal = {Physics of Atomic Nuclei},
publisher = {Pleiades Publishing Ltd},
year = {2022},
month = {April},
pages = {167–175}
}

@article{10.1140/epjc/s10052-019-6863-8,
author = {{CMS Collaboration} and others},
title = {Measurement of the $t\overline{t}$ production cross section, the top quark mass, and the strong coupling constant using dilepton events in pp collisions at $\sqrt{s}$ = 13 TeV},
journal = {The European Physical Journal C},
volume = {79},
pages = {1-36},
year = {2019},
month = {April},
day = {29},
number = {5},
publisher = {Springer Science and Business Media LLC},
issn = {1434-6052},
doi = {10.1140/epjc/s10052-019-6863-8},
url = {https://dx.doi.org/10.1140/epjc/s10052-019-6863-8}
}

@article{10.1016/j.physletb.2018.03.023,
author = {{Atlas Collaboration and others}},
title = {Measurement of the production cross-section of a single top quark in association with a {Z} boson in proton–proton collisions at 13 {TeV} with the {ATLAS} detector},
volume = {780},
issn = {0370-2693},
doi = {10.1016/j.physletb.2018.03.023},
url = {https://dx.doi.org/10.1016/j.physletb.2018.03.023},
journal = {Physics Letters B},
publisher = {Elsevier BV},
year = {2018},
month = {May},
pages = {557–577}
}

@article{10.1140/epjc/s10052-011-1554-0,
   title={Asymptotic formulae for likelihood-based tests of new physics},
   volume={71},
   ISSN={1434-6052},
   url={http://dx.doi.org/10.1140/epjc/s10052-011-1554-0},
   DOI={10.1140/epjc/s10052-011-1554-0},
   number={2},
   pages={1554},
   journal={The European Physical Journal C},
   publisher={Springer Science and Business Media LLC},
   author={Cowan, Glen and Cranmer, Kyle and Gross, Eilam and Vitells, Ofer},
   year={2011},
   month=feb
}

@misc{2203.11601,
    author = {Christian Bierlich and Smita Chakraborty and Nishita Desai and Leif Gellersen and Ilkka Helenius and Philip Ilten and Leif Lönnblad and Stephen Mrenna and Stefan Prestel and Christian T. Preuss and Torbjörn Sjöstrand and Peter Skands and Marius Utheim and Rob Verheyen},
    title = {A comprehensive guide to the physics and usage of PYTHIA 8.3},
    year = {2022},
    journal={SciPost Physics Codebases},
    pages={008},
    eprint={2203.11601},
    archivePrefix={arXiv},
    primaryClass={hep-ph},
    url={https://arxiv.org/abs/2203.11601}, 
    langid = {english},
    language = {english},
    hyphenation={english}
}

@article{10.1140/epjc/s10052-016-4018-8,
   title={Herwig 7.0/Herwig++ 3.0 release note},
   volume={76},
   ISSN={1434-6052},
   url={http://dx.doi.org/10.1140/epjc/s10052-016-4018-8},
   DOI={10.1140/epjc/s10052-016-4018-8},
   number={4},
   journal={The European Physical Journal C},
   publisher={Springer Science and Business Media LLC},
   author={Bellm, Johannes and Gieseke, Stefan and Grellscheid, David and Pl\"{a}tzer, Simon and Rauch, Michael and Reuschle, Christian and Richardson, Peter and Schichtel, Peter and Seymour, Michael H. and Siódmok, Andrzej and Wilcock, Alexandra and Fischer, Nadine and Harrendorf, Marco A. and Nail, Graeme and Papaefstathiou, Andreas and Rauch, Daniel},
   year={2016},
   month={April}
}

@article{10.1140/epjc/s10052-012-1896-2,
    author = {Cacciari, Matteo and Salam, Gavin P. and Soyez, Gregory},
    title = {FastJet user manual},
    journal = {The European Physical Journal C},
    year = {2012},
    volume = {72},
    ISSN={1434-6052},
    url={http://dx.doi.org/10.1140/epjc/s10052-012-1896-2},
    DOI={10.1140/epjc/s10052-012-1896-2},
    number={3},
    publisher={Springer Science and Business Media LLC},
    month={March},
    langid = {english},
    language = {english},
    hyphenation={english}
}

@article{10.1140/epjc/s10052-017-4852-3,
  doi = {10.1140/epjc/s10052-017-4852-3},
  url = {http://dx.doi.org/10.1140/epjc/s10052-017-4852-3},
  author = {{ATLAS Collaboration}},
  title = {{Performance of the ATLAS Trigger System in 2015}},
  volume={77},
  ISSN={1434-6052},
  journal={The European Physical Journal C},
  publisher={Springer Science and Business Media LLC},
  year = {2017}
}

@article{10.1088/1126-6708/2008/04/063,
    author = {Cacciari, Matteo and Salam, Gavin P and Soyez, Gregory},
    title = {The anti-k\_t jet clustering algorithm},
    journal = {Journal of High Energy Physics},
    publisher={Springer Science and Business Media LLC},
    year = {2008},
    volume = {2008},
    ISSN={1029-8479},
    number={04},
    url={http://dx.doi.org/10.1088/1126-6708/2008/04/063},
    DOI={10.1088/1126-6708/2008/04/063},
    month={April},
    pages={063–063},
    langid = {english},
    language = {english},
    hyphenation={english}
}

@article{10.1088/1748-0221/11/04/p04008,
author = {{Atlas Collaboration and others}},
title = {{Performance of b-jet identification in the ATLAS experiment}},
volume = {11},
issn = {1748-0221},
url = {https://dx.doi.org/10.1088/1748-0221/11/04/P04008},
doi = {10.1088/1748-0221/11/04/p04008},
number = {04},
journal = {Journal of Instrumentation},
publisher = {IOP Publishing},
year = {2016},
month = {April},
pages = {P04008–P04008},
langid = {english},
language = {english},
hyphenation={english}
}

@article{10.1088/1748-0221/13/05/p05011,
author={{CMS Collaboration and others}},
title = {{Identification of heavy-flavour jets with the CMS detector in pp collisions at 13 TeV}},
volume = {13},
issn = {1748-0221},
url = {https://dx.doi.org/10.1088/1748-0221/13/05/P05011},
doi = {10.1088/1748-0221/13/05/p05011},
number = {05},
journal = {Journal of Instrumentation},
publisher = {IOP Publishing},
year = {2018},
month = {May},
pages = {P05011–P05011},
langid = {english},
language = {english},
hyphenation={english}
}

@book{10.1142/12294,
  title = {Deep Learning for Physics Research},
  ISBN = {9789811237461},
  url = {http://dx.doi.org/10.1142/12294},
  DOI = {10.1142/12294},
  publisher = {WORLD SCIENTIFIC},
  author = {Erdmann,  Martin and Glombitza,  Jonas and Kasieczka,  Gregor and Klemradt,  Uwe},
  year = {2021},
  month = {February} 
}

@book{Goodfellow-et-al-2016,
    title={Deep Learning},
    author={Ian Goodfellow and Yoshua Bengio and Aaron Courville},
    publisher={MIT Press},
    note={\url{http://www.deeplearningbook.org}},
    year={2016}
}

@article{10.1038/nature14539,
  title = {Deep learning},
  volume = {521},
  ISSN = {1476-4687},
  url = {http://dx.doi.org/10.1038/nature14539},
  DOI = {10.1038/nature14539},
  number = {7553},
  journal = {Nature},
  publisher = {Springer Science and Business Media LLC},
  author = {LeCun,  Yann and Bengio,  Yoshua and Hinton,  Geoffrey},
  year = {2015},
  month = {May},
  pages = {436–444}
}

@article{10.1007/bf02551274,
  title = {Approximation by superpositions of a sigmoidal function},
  volume = {2},
  ISSN = {1435-568X},
  url = {http://dx.doi.org/10.1007/bf02551274},
  DOI = {10.1007/bf02551274},
  number = {4},
  journal = {Mathematics of Control,  Signals,  and Systems},
  publisher = {Springer Science and Business Media LLC},
  author = {Cybenko,  G.},
  year = {1989},
  month = {December},
  pages = {303–314}
}

@article{10.1016/j.heliyon.2024.e26586,
  title = {{Improving classification accuracy of fine-tuned CNN models: Impact of hyperparameter optimization}},
  volume = {10},
  ISSN = {2405-8440},
  url = {http://dx.doi.org/10.1016/j.heliyon.2024.e26586},
  DOI = {10.1016/j.heliyon.2024.e26586},
  number = {5},
  journal = {Heliyon},
  publisher = {Elsevier BV},
  author = {Wojciuk,  Mikolaj and Swiderska-Chadaj,  Zaneta and Siwek,  Krzysztof and Gertych,  Arkadiusz},
  year = {2024},
  month = {March},
  pages = {e26586}
}

@misc{frazier2018tutorialbayesianoptimization,
      title={A Tutorial on Bayesian Optimization}, 
      author={Peter I. Frazier},
      year={2018},
      eprint={1807.02811},
      archivePrefix={arXiv},
      primaryClass={stat.ML},
      url={https://arxiv.org/abs/1807.02811}, 
}

@misc{10.48550/ARXIV.1801.01596,
  doi = {10.48550/ARXIV.1801.01596},
  url = {https://arxiv.org/abs/1801.01596},
  author = {Wang,  Jiazhuo and Xu,  Jason and Wang,  Xuejun},
  title = {Combination of Hyperband and Bayesian Optimization for Hyperparameter Optimization in Deep Learning},
  publisher = {arXiv},
  year = {2018},
  eprint={1801.01596},
  archivePrefix={arXiv},
  copyright = {arXiv.org perpetual,  non-exclusive license}
}

@article{10.11989/JEST.1674-862X.80904120,
title = {Hyperparameter Optimization for Machine Learning Models Based on Bayesian Optimizationb},
journal = {Journal of Electronic Science and Technology},
volume = {17},
number = {1},
pages = {26-40},
year = {2019},
issn = {1674-862X},
doi = {10.11989/JEST.1674-862X.80904120},
url = {https://doi.org/10.11989/JEST.1674-862X.80904120},
author = {Jia Wu and Xiu-Yun Chen and Hao Zhang and Li-Dong Xiong and Hang Lei and Si-Hao Deng}
}

@misc{10.48550/arxiv.2304.11127,
  doi = {10.48550/arxiv.2304.11127},
  url = {https://doi.org/10.48550/arxiv.2304.11127},
  author = {Watanabe,  Shuhei},
  title = {Tree-Structured Parzen Estimator: Understanding Its Algorithm Components and Their Roles for Better Empirical Performance},
  publisher = {arXiv},
  year = {2023},
  copyright = {arXiv.org perpetual,  non-exclusive license}
}

@inproceedings{10.1145/3292500.3330701,
  author    = {Akiba, Takuya and Sano, Shotaro and Yanase, Toshihiko and Ohta, Takeru and Koyama, Masanori},
  title     = {Optuna: A Next-generation Hyperparameter Optimization Framework},
  booktitle = {Proceedings of the 25th ACM SIGKDD International Conference on Knowledge Discovery \& Data Mining},
  year      = {2019},
  month     = jul,
  pages     = {2623--2631},
  publisher = {ACM},
  doi       = {10.1145/3292500.3330701},
  url       = {https://doi.org/10.1145/3292500.3330701}
}

@article{10.1016/j.neucom.2020.07.061,
  title = {On hyperparameter optimization of machine learning algorithms: Theory and practice},
  volume = {415},
  ISSN = {0925--2312},
  url = {http://dx.doi.org/10.1016/j.neucom.2020.07.061},
  DOI = {10.1016/j.neucom.2020.07.061},
  journal = {Neurocomputing},
  publisher = {Elsevier BV},
  author = {Yang,  Li and Shami,  Abdallah},
  year = {2020},
  month = Nov,
  pages = {295–316}
}

@article{10.1002/widm.1484,
  title = {Hyperparameter optimization: Foundations,  algorithms,  best practices,  and open challenges},
  volume = {13},
  ISSN = {1942--4795},
  url = {http://dx.doi.org/10.1002/widm.1484},
  DOI = {10.1002/widm.1484},
  number = {2},
  journal = {WIREs Data Mining and Knowledge Discovery},
  publisher = {Wiley},
  author = {Bischl,  Bernd and Binder,  Martin and Lang,  Michel and Pielok,  Tobias and Richter,  Jakob and Coors,  Stefan and Thomas,  Janek and Ullmann,  Theresa and Becker,  Marc and Boulesteix,  Anne‐Laure and Deng,  Difan and Lindauer,  Marius},
  year = {2023},
  month = Jan 
}

@article{10.1088/1748-0221/19/11/p11021,
  title = {Performance of the CMS high-level trigger during LHC Run 2},
  volume = {19},
  ISSN = {1748-0221},
  url = {http://dx.doi.org/10.1088/1748-0221/19/11/p11021},
  DOI = {10.1088/1748-0221/19/11/p11021},
  number = {11},
  journal = {Journal of Instrumentation},
  publisher = {IOP Publishing},
  author = {{CMS Collaboration} and others},
  year = {2024},
  month = Nov,
  pages = {P11021}
}

@article{10.1088/1748-0221/12/01/p01020,
  title = {The CMS trigger system},
  volume = {12},
  ISSN = {1748-0221},
  url = {http://dx.doi.org/10.1088/1748-0221/12/01/p01020},
  DOI = {10.1088/1748-0221/12/01/p01020},
  number = {01},
  journal = {Journal of Instrumentation},
  publisher = {IOP Publishing},
  author = {{CMS Collaboration} and others},
  year = {2017},
  month = Jan,
  pages = {P01020–P01020}
}

@article{10.1088/1748-0221/19/05/p05064,
  title = {Development of the CMS detector for the CERN LHC Run 3},
  volume = {19},
  ISSN = {1748-0221},
  url = {http://dx.doi.org/10.1088/1748-0221/19/05/p05064},
  DOI = {10.1088/1748-0221/19/05/p05064},
  number = {05},
  journal = {Journal of Instrumentation},
  publisher = {IOP Publishing},
  author = {{CMS Collaboration and others}},
  year = {2024},
  month = May,
  pages = {P05064}
}

@article{10.1088/1748-0221/19/06/p06029,
  title = {{The ATLAS trigger system for LHC Run 3 and trigger performance in 2022}},
  volume = {19},
  ISSN = {1748-0221},
  url = {http://dx.doi.org/10.1088/1748-0221/19/06/p06029},
  DOI = {10.1088/1748-0221/19/06/p06029},
  number = {06},
  journal = {Journal of Instrumentation},
  publisher = {IOP Publishing},
  author = {{ATLAS Collaboration and others}},
  year = {2024},
  month = jun,
  pages = {P06029}
}

@article{10.1213/ane.0000000000002864,
  title = {Correlation Coefficients: Appropriate Use and Interpretation},
  volume = {126},
  ISSN = {0003-2999},
  url = {http://dx.doi.org/10.1213/ANE.0000000000002864},
  DOI = {10.1213/ane.0000000000002864},
  number = {5},
  journal = {Anesthesia \& Analgesia},
  publisher = {Ovid Technologies (Wolters Kluwer Health)},
  author = {Schober,  Patrick and Boer,  Christa and Schwarte,  Lothar A.},
  year = {2018},
  month = May,
  pages = {1763–1768}
}

@article{10.1098/rspl.1895.0041,
  title = {VII. Note on regression and inheritance in the case of two parents},
  volume = {58},
  ISSN = {2053-9126},
  url = {http://dx.doi.org/10.1098/rspl.1895.0041},
  DOI = {10.1098/rspl.1895.0041},
  number = {347-352},
  journal = {Proceedings of the Royal Society of London},
  publisher = {The Royal Society},
  author = {Pearson,  Karl},
  year = {1895},
  month = Dec,
  pages = {240–242}
}

@article{10.1023/a:1010933404324,
  title = {Random Forests},
  volume = {45},
  ISSN = {1573-0565},
  url = {http://dx.doi.org/10.1023/a:1010933404324},
  DOI = {10.1023/a:1010933404324},
  number = {1},
  journal = {Machine Learning},
  publisher = {Springer Science and Business Media LLC},
  author = {Breiman,  Leo},
  year = {2001},
  month = Oct,
  pages = {5–32}
}

@article{10.1016/j.physletb.2018.07.035,
  title = {Observation of Higgs boson production in association with a top quark pair at the LHC with the ATLAS detector},
  volume = {784},
  ISSN = {0370-2693},
  url = {http://dx.doi.org/10.1016/j.physletb.2018.07.035},
  DOI = {10.1016/j.physletb.2018.07.035},
  journal = {Physics Letters B},
  publisher = {Elsevier BV},
  author = {{Atlas Collaboration and others}},
  year = {2018},
  month = Sep,
  pages = {173–191}
}

@article{10.1103/physrevlett.131.231901,
  title = {Precise Predictions for the Associated Production of a W Boson with a Top-Antitop Quark Pair at the LHC},
  volume = {131},
  ISSN = {1079-7114},
  url = {http://dx.doi.org/10.1103/PhysRevLett.131.231901},
  DOI = {10.1103/physrevlett.131.231901},
  number = {23},
  journal = {Physical Review Letters},
  publisher = {American Physical Society (APS)},
  author = {Buonocore,  Luca and Devoto,  Simone and Grazzini,  Massimiliano and Kallweit,  Stefan and Mazzitelli,  Javier and Rottoli,  Luca and Savoini,  Chiara},
  year = {2023},
  month = Dec 
}

@article{10.3390/universe9050242,
  title = {Measurements of the Cross-Section for the $t\bar{t}b\bar{b}$ Heavy-Flavor Production at the LHC},
  volume = {9},
  ISSN = {2218-1997},
  url = {http://dx.doi.org/10.3390/universe9050242},
  DOI = {10.3390/universe9050242},
  number = {5},
  journal = {Universe},
  publisher = {MDPI AG},
  author = {D’Hondt,  Jorgen and Kim,  Tae},
  year = {2023},
  month = May,
  pages = {242}
}

@article{10.1162/neco.1991.3.4.461,
  title = {Neural Network Classifiers Estimate Bayesian a posteriori Probabilities},
  volume = {3},
  ISSN = {1530-888X},
  url = {http://dx.doi.org/10.1162/neco.1991.3.4.461},
  DOI = {10.1162/neco.1991.3.4.461},
  number = {4},
  journal = {Neural Computation},
  publisher = {MIT Press - Journals},
  author = {Richard,  Michael D. and Lippmann,  Richard P.},
  year = {1991},
  month = dec,
  pages = {461–483}
}

@article{10.1098/rsta.1933.0009,
  title = {IX. On the problem of the most efficient tests of statistical hypotheses},
  volume = {231},
  ISSN = {2053-9258},
  url = {http://dx.doi.org/10.1098/rsta.1933.0009},
  DOI = {10.1098/rsta.1933.0009},
  number = {694-706},
  journal = {Philosophical Transactions of the Royal Society of London. Series A,  Containing Papers of a Mathematical or Physical Character},
  publisher = {The Royal Society},
  author = {Neyman,  Jerzy and Pearson,  Egon Sharpe},
  year = {1933},
  month = feb,
  pages = {289–337}
}

@article{10.1038/s41586-022-04893-w,
  title = {A detailed map of Higgs boson interactions by the ATLAS experiment ten years after the discovery},
  volume = {607},
  ISSN = {1476-4687},
  url = {http://dx.doi.org/10.1038/s41586-022-04893-w},
  DOI = {10.1038/s41586-022-04893-w},
  number = {7917},
  journal = {Nature},
  publisher = {Springer Science and Business Media LLC},
  author = {{Atlas Collaboration and others}},
  year = {2022},
  month = jul,
  pages = {52–59}
}

@article{10.1140/epjc/s10052-019-6909-y,
  title = {Combined measurements of Higgs boson couplings in proton–proton collisions at $\sqrt{s}=13\, \text {Te}\text {V}$},
  volume = {79},
  ISSN = {1434-6052},
  url = {http://dx.doi.org/10.1140/epjc/s10052-019-6909-y},
  DOI = {10.1140/epjc/s10052-019-6909-y},
  number = {5},
  journal = {The European Physical Journal C},
  publisher = {Springer Science and Business Media LLC},
  author={{CMS Collaboration and others}},
  year = {2019},
  month = May 
}

@article{10.1088/1748-0221/14/08/P08020,
  title = {Machine Learning on data with sPlot background subtraction},
  volume = {14},
  issn = {1748-0221},
  url = {https://dx.doi.org/10.1088/1748-0221/14/08/P08020},
  doi = {10.1088/1748-0221/14/08/p08020},
  number = {08},
  journal = {Journal of Instrumentation},
  publisher = {IOP Publishing},
  author = {Borisyak,  M. and Kazeev,  N.},
  year = {2019},
  month = Aug,
  pages = {P08020–P08020}
}

@article{10.48550/arxiv.physics/0703039,
  doi = {10.48550/arxiv.physics/0703039},
  url = {https://doi.org/10.48550/arxiv.physics/0703039},
  author = {Hoecker,  A. and Speckmayer,  P. and Stelzer,  J. and Therhaag,  J. and von Toerne,  E. and Voss,  H. and Backes,  M. and Carli,  T. and Cohen,  O. and Christov,  A. and Dannheim,  D. and Danielowski,  K. and Henrot-Versille,  S. and Jachowski,  M. and Kraszewski,  K. and Krasznahorkay,  A. and Kruk,  M. and Mahalalel,  Y. and Ospanov,  R. and Prudent,  X. and Robert,  A. and Schouten,  D. and Tegenfeldt,  F. and Voigt,  A. and Voss,  K. and Wolter,  M. and Zemla,  A.},
  title = {{TMVA} -- Toolkit for Multivariate Data Analysis},
  year = {2007},
  pages={040}
}

@article{10.1214/aos/1176344552,
  title = {Bootstrap Methods: Another Look at the Jackknife},
  volume = {7},
  issn = {2168-8966},
  url = {http://dx.doi.org/10.1214/aos/1176344552},
  doi = {10.1214/aos/1176344552},
  number = {1},
  journal = {The Annals of Statistics},
  publisher = {Institute of Mathematical Statistics},
  author = {Efron,  B.},
  year = {1979},
  month = Jan,
  pages = {1–26}
}

\end{document}